\documentclass[a4paper,12pt]{elsarticle}

\usepackage{framed,multirow}
\usepackage{amssymb,amsmath}
\usepackage{latexsym}
\usepackage{url}
\usepackage{xcolor}
\usepackage{bm}
\usepackage{amsmath}
\usepackage{algorithm}
\usepackage{algpseudocode}
\usepackage{mathtools}
\usepackage{lineno}
\usepackage[thinlines]{easytable}
\usepackage{todonotes}

\usepackage[margin=2.5cm]{geometry}

\newcommand{\p}{\partial}
\newcommand{\D}{\text{d}}

\graphicspath{{Figures/}}

\begin{document}


\begin{frontmatter}

\title{Effects of different discretisations of the Laplacian upon stochastic
simulations of reaction-diffusion systems \\ on both static and growing domains}%

\author{Bartosz J. Bartmanski$^\dagger$ and Ruth E. Baker$^\dagger$}

\address{$^\dagger$Mathematical Institute, University of Oxford, \\ Woodstock Road, Oxford, OX2 6GG, UK}


\begin{abstract}
By discretising space into compartments and letting system dynamics be governed by the reaction-diffusion master equation, it is possible to derive and simulate a stochastic model of reaction and diffusion on an arbitrary domain. However, there are many implementation choices involved in this process, such as the choice of discretisation and method of derivation of the diffusive jump rates, and it is not clear \textit{a priori} how these affect model predictions. To shed light on this issue, in this work we explore how a variety of discretisations and method for derivation of the diffusive jump rates affect the outputs of stochastic simulations of reaction-diffusion models, in particular using Turing's model of pattern formation as a key example. We consider both static and uniformly growing domains and demonstrate that, while only minor differences are observed for simple reaction-diffusion systems, there can be vast differences in model predictions for systems that include complicated reaction kinetics, such as Turing's model of pattern formation. Our work highlights that care must be taken in using the reaction-diffusion master equation to make predictions as to the dynamics of stochastic reaction-diffusion systems.
\end{abstract}




\end{frontmatter}



\pagebreak


\section{Introduction}
\label{Sec:introduction}

Many biological processes hinge on the reaction and diffusion of constituent components and most are, in addition, subject to the inherent stochasticity of the natural world. Due to the nonlinearity of interacting components,  mathematical models provide a crucial means to interrogate how the combination of reaction and diffusion drive a given process~\cite{Elowitz2002,Howard2003,Isaacson2011,Longo2006,Mort2016,Volkening2015}.


Models of reaction-diffusion systems can be roughly divided into macroscopic, mesoscopic and microscopic~\cite{Hellander2017}. On the macroscale, partial differential equations (PDEs) are often used to model reaction-diffusion systems in terms of the evolution of species concentration in time and space. Such models can be analysed using a range of existing mathematical tools, however they do not take into account stochastic effects. At the other end of the spectrum we have microscopic descriptions such as molecular dynamics models, where each individual molecule is described by its position and velocity, and molecules react when within a certain radius of each other~\cite{anderson2007modified}. Although they provide a very detailed account of the reaction-diffusion process under consideration, they incur extremely high computational cost and are, as such, unfeasible for many real-world models~\cite{Osborne2017,VanLiedekerke2015}. A mesoscopic description aims to strike a balance between the two, including descriptions of individual molecule dynamics without incurring prohibitive computational overheads. An example mesoscopic description is that of the reaction-diffusion master equation framework (RDME)~\cite{Isaacson2009,Isaacson2013,vanKampen2007}; in this model, the domain is discretised into compartments, molecules can react with other molecules in the same compartment, and diffusion is represented by jumps between neighbouring compartments. The RDME framework has been used to explore a range of biological phenomena, and it is this description of  reaction and diffusion on which we focus in this work.

To model diffusion, the RDME framework requires computation of the rate per unit time of a molecule
to jump from one compartment to a neighbouring one. Previous research in this direction includes the work
by Engblom \textit{et al.}~\cite{engblom2009simulation}, where the finite element method is used to calculate the diffusive jump rates, and that of Hellander and Petzold~\cite{Hellander2017} who developed a description of diffusion for a more general compartment shapes. In addition, Meinecke and L\"otstedt explored the impact of different methods of diffusive jump rate derivation upon simulation of diffusive systems~\cite{Lotstedt2015,meinecke2016stochastic,Meinecke2016}. To mitigate issues with convergence of the RDME in the context of higher order reactions, Isaacson developed a convergent RDME that allows molecules to react even when they are not in the same compartment~\cite{Isaacson2013}. However, this approach it does not consider how differences in the derivation of the diffusive jump rates affect model predictions.


\subsection{Aims and outline}

Our aim in this work is to understand how the different methods of derivation of diffusive jump rates, and different domain discretisations, affect model predictions for reaction-diffusion systems, including those with bimolecular reactions and reaction kinetics capable of pattern formation. We will use Meinecke and L\"otstedt's~\cite{meinecke2016stochastic} derivations of diffusive jump rates as a basis for our work. These derivations are based on three different numerical schemes: the finite volume method (FVM); the finite element method (FEM); and the finite difference method (FDM). In addition we include the fourth approach taken by Meinecke and L{\"o}tstedt that is based on first exit times (FETs)~\cite{meinecke2016stochastic}. In Section~\ref{Sec:model} we outline the results of Meinecke and L{\"o}tstedt for a static domain~\cite{meinecke2016stochastic}, extend them to a uniformly growing domain, and present the stochastic simulation algorithm that we will use throughout this work. Our results are outlined in Section~\ref{Sec:results}, including the effects of different diffusive jump rates on pattern formation in Section~\ref{Sec:results_schnakenberg_kinetics}. A brief discussion of the results outlined in this paper can be found Section~\ref{Sec:conclusions}.


\section{The reaction-diffusion master equation framework}
\label{Sec:model}

The chemical master equation (CME) framework describes the reaction of chemical species under the assumption that the molecules are well-mixed~\cite{Schnoerr2017}. In general, the CME is intractable and, as a result, Gillespie~\cite{gillespie1976general} developed a stochastic simulation algorithm (SSA) to generate sample paths of the system under consideration. The RDME framework is an extension of the CME where the spatial domain is discretised into compartments, each of which is assumed well-mixed, and diffusion is modelled using a
set of first-order reactions that describe the jumps of molecules between neighbouring compartments~\cite{Schnoerr2017}. Again, the RDME is rarely tractable, and so stochastic simulation algorithms have been developed to simulate reaction-diffusion systems modelled using the RDME framework~\cite{Elf2004,Hattne2005}.

In this section, we outline the RDME framework, and the stochastic simulation algorithm that we will use throughout this work. For simplicity, we consider the evolution of a reaction-diffusion system in the square domain $\Omega = [0, L]\times[0, L]$, and partition $\Omega$ using a Cartesian mesh into non-overlapping rectangular sub-domains $C_i$, called compartments, such that $\Omega = \bigcup_{i=1}^{I} C_i$ (Figure~\ref{Fig:diagram})~\cite{meinecke2016stochastic}. Each compartment is of size $\kappa h \times h$, where $\kappa$ is the aspect ratio of a compartment, and the total number of compartments, $I$, can be expressed as $I=n_x \times n_y$, where $n_x$ and $n_y$ are the numbers of compartments in the $x$-direction and $y$-direction, respectively. We number the compartments starting at the bottom-left corner of the domain and then from left to right, so that, for site $i$, the adjacent compartments have indices $i\pm 1$ in the $x$-direction and indices $i\pm n_x$ in the $y$-direction.

We consider a reaction-diffusion system consisting of $L$ species, where $U_i^\ell(t)$ is the number of molecules of species $\ell$ in compartment $i$ at time $t$, for $\ell=1,\ldots,L$ and $i=1,\ldots,I$. We define $\bm{U}^\ell(t)=(U_1^\ell(t),\ldots,U^\ell_I(t))$ to be the spatial vector of species $\ell$ at time $t$, $\bm{U}(t)=(\bm{U}^1(t),\ldots,\bm{U}^L(t))$ to be the state matrix of the system and $\mathbb{P}(\bm{u},t)=\mathbb{P}(\bm{U}(t)=\bm{u}|\bm{U}(0)=\bm{u}_0)$. Changes to the state matrix occur through either reactions or diffusive jumps. We consider a system of $r=1,\ldots,R$ chemical reactions such that reaction $r$ has propensity function $a_i^r$ in compartment $i$ and is defined such that
\begin{eqnarray}
\nonumber
a_i^r(\bm{u}(t))\D{t}&\coloneqq&\mathbb{P}\left(\text{reaction $r$ fires in compartment $i$ during $[t,t+\D{t})$} \right. \\ && \left.\qquad\quad\text{given the system is in state $\bm{u}$ at time $t$}\right),
\end{eqnarray}
where $\D{t}$ is an infinitesimal time interval. For more details of the specific forms of the propensity functions see~\cite{Schnoerr2017}. Similarly, diffusive jumps have propensity function $d_{i,j}^\ell$ where
\begin{equation}
d_{i,j}^{\,\ell}(\bm{u}(t))\D{t}=\lambda^\ell_{i,j}u_i^\ell(t)\D{t},
\end{equation}
is the probability that a molecule of species $\ell$ jumps from compartment $i$ to compartment $j$ in the infinitesimal time interval $[t,t+\D{t})$ given the system is in state $\bm{u}$ at time $t$.

As such, the RDME can be written as
\begin{eqnarray}
\nonumber
\frac{\p}{\p{t}}\mathbb{P}\left(\bm{u},t\right)&=&
\sum_{i=1}^I \sum_{r=1}^R \left[a_i^r\left(\textbf{u}-\bm{\nu}_{i,r}\right)\mathbb{P}\left(\bm{u}-\bm{\nu}_{i,r},t\right)-a_i^r\left(\textbf{u}\right)\mathbb{P}\left(\bm{u},t\right)\right] \\
&& \quad +
\sum_{i=1}^I\sum_{\substack{j=1\\j\neq{i}}}^I\sum_{\ell=1}^L \left[ \lambda^\ell_{i,j}\left(u_i^\ell(t)+1\right)\mathbb{P}\left(\bm{u}-\bm{e}_{i,j}^\ell,t\right) - \lambda^\ell_{i,j}u_i^\ell(t)\mathbb{P}\left(\bm{u},t\right)\right],
\label{equation:RDME}
\end{eqnarray}
where $\bm{\nu}_{i,r}$ is the stochiometric matrix that describes the change in state upon firing of reaction $r$ in compartment $i$, and $\bm{e}_{i,j}^\ell$ is the stochiometric matrix that describes the change in state upon a diffusive jump of a molecule of species $\ell$ from compartment $i$ to compartment $j$~\cite{Isaacson2009}.

To generate sample paths for a reaction-diffusion system within the RDME framework described in Equation~\eqref{equation:RDME}, we use an SSA called the next sub-volume method~\cite{Elf2004,Hattne2005}. Let $t_i$ be the time until the next reaction within compartment $i$ for $i=1,\ldots,I$, and let the total propensity for compartment $i$ be
\begin{equation}
a_i^0(\bm{u}(t))=\sum_{k=1}^K a_i^k(\bm{u}(t))+\sum_{\substack{j=1\\j\neq{i}}}^I\sum_{\ell=1}^L{d_{i,j}^{\,\ell}(\bm{u}(t))}.
\label{Eq:total_propensity}
\end{equation}
Then, given that $T$ is the end time of a simulation, we can simulate sample paths consistent with the RDME in Equation~\eqref{equation:RDME} using the SSA given in Algorithm~\ref{Alg:SSA}.


\begin{algorithm}
	\caption{\textmd{Next sub-volume method SSA \label{Alg:SSA}}}
\vspace{1ex}
\begin{algorithmic}
	\State Set $t=0$.
	\State Initialise $\bm{u}$.
  \For{$i$ in $\left\{1,2,\ldots,I\right\}$}
    \State Initialise the time until the next reaction for compartment $i$: $t_i \sim \exp(a_i^0(\bm{u}(t)))$.
  \EndFor
	\While{$t < T$}
    \State Find argmin of the set $\left\{t_1,t_2,\ldots,t_I\right\}$ and denote it $m$.
    \State Set $t = t_m$.
    \State Generate a random number $r \sim \textit{U}(0,1)$.
    \State Choose reaction / diffusion $q$ to fire. Note that reaction $k$ fires with probability $a_i^k(\bm{u}(t))/a_i^0(\bm{u}(t))$ and diffusion of a molecule of species $\ell$ from box $i$ to box $j$ occurs with probability $d_{i,j}^{\,\ell}(\bm{u}(t))/a_i^0(\bm{u}(t))$.
	\State Update molecule numbers: $\bm{u} := \bm{u} + \bm{\nu}_{q}$, where $\bm{\nu}_{q}$ is the stoichiometric matrix of reaction / diffusion $q$.
    \State Let $\Gamma$ be the set of indices of non-zero elements of $\bm{\nu}_q$.
    \For{$\gamma$ in $\Gamma$}
      \State Update $a_i^k(\bm{u}(t))$, $d_{i,j}^{\,\ell}(\bm{u}(t))$ and $a_i^0(\bm{u})$ accordingly.
      \State Generate a random number $\Delta t \sim \exp(a_\gamma^0(\bm{u}))$.
      \State Update the time until the next reaction: $t_{\gamma} = t + \Delta t$.
    \EndFor
	\EndWhile
\end{algorithmic}
\vspace{1ex}
\end{algorithm}


Although the RDME as outlined in Equation~\eqref{equation:RDME} and the accompanying text, together with the SSA of Algorithm~\ref{Alg:SSA}, holds in full generality, for the rest of this work, we will assume for simplicity that all diffusion coefficients and reaction rates are spatially homogeneous.



\subsection{Derivations of the diffusive jump rates on a static domain}
\label{Sec:diffusion}

We first consider a diffusion-only system with a single molecular species, U. As already outlined in Section~\ref{Sec:model}, diffusion is modelled as a series of jumps between neighbouring compartments where the propensity function for a jump from compartment $i$ to compartment $j$ is $d_{i,j}(\bm{u}(t))=\lambda_{i,j}u_i(t)$. Note that we have dropped the superscript $\ell$ for clarity of notation since we are currently considering only a single species. In this work we will assume that $\lambda_{i,j}=0$, $i=1,\ldots,I$, unless $j$ is one of the eight nearest neighbour compartments (indices $i\pm1$, $i\pm{n_x}$ and $i\pm(n_x\pm1)$), as depicted in Figure~\ref{Fig:diagram}. In this section we outline four methods for derivation of the $\lambda_{i,j}$.


\subsubsection{Diffusion of a single molecule}

Consider a single molecule diffusing within the domain $\Omega=[0, L]\times[0, L]$ with diffusion coefficient $D$. Then evolution of the position, $\textbf{x}(t)=(x(t),y(t))$, of this molecule can be described using the following stochastic differential equations (SDEs)~\cite{Erban2007}:
\begin{eqnarray}
x(t+\D{t})&=&x(t)+\sqrt{2D}\D{W_x},\\
y(t+\D{t})&=&y(t)+\sqrt{2D}\D{W}_y,
\end{eqnarray}
where $W_x$ and $W_y$ are the usual Wiener processes. The associated Fokker-Planck equation (FPE) is
\begin{equation}
\frac{\p p(\textbf{x}, t)}{\p t} = D \Delta p (\textbf{x}, t)
\quad \text{for} \quad \textbf{x} \in \Omega,
\label{Eq:diff_eq}
\end{equation}
so that $p(\textbf{x},t)\D{\textbf{x}}$ is the probability to find the molecule in the region $[x,x+\D{x}]\times[y,y+\D{y}]$ at time $t$.

For simplicity, throughout this work we will use Neumann boundary conditions
\begin{equation}
\left( \textbf{n} \cdot \nabla \right) p = 0 \quad
\text{for} \quad \textbf{x} \in \partial \Omega,
\label{Eq:diff_Eq:bc}
\end{equation}
and we take the initial distribution for the molecule to be
\begin{equation}
p(\textbf{x}, 0) = p_0 (\textbf{x} ).
\label{Eq:diff_Eq:ic}
\end{equation}

\begin{figure}
\centering
\includegraphics[scale=1.0]{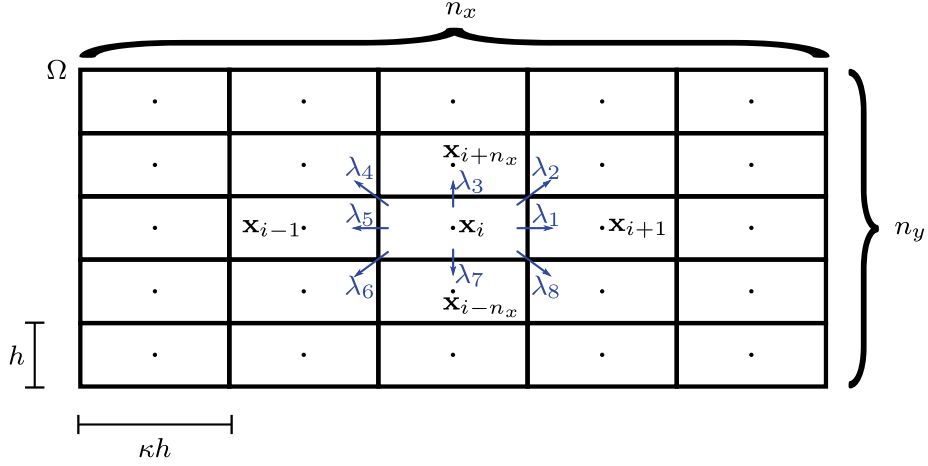}
\caption{Schematic diagram of the discretisation of the two-dimensional domain
$\Omega$, with the dimensions of the compartments and the indexing of the jump rates,
$\lambda_n$, for $n \in \left\lbrace 1, \ldots, 8 \right\rbrace$
for a given compartment $i$ shown.}
\label{Fig:diagram}
\end{figure}

Using the Cartesian mesh shown in Figure~\ref{Fig:diagram}, we discretise the Laplacian as follows:
\begin{equation}
D \Delta p ( \textbf{x}_i, t )\approx
\sum_{j=1}^{8} \lambda_{i,j} \, \tilde{p}_j ( t )- \sum_{j=1}^{8} \lambda_{j,i} \, \tilde{p}_i ( t ),
\label{Eq:discretised_laplacian}
\end{equation}
where $\tilde{p}_i(t)$ is the approximate solution to Equation~\eqref{Eq:diff_eq} at point $\textbf{x}_i$, i.e. at the centre of compartment $i$ at time $t$. The compartments are all of the same size, $\kappa h \times h$, with area $A = \kappa h^2$ (Figure~\ref{Fig:diagram}). Hence, exploiting translational symmetry and noting that the diffusion coefficient, $D$, is spatially homogeneous, we refer to the jump rates, $\lambda_{i,j}$, from now on as $\lambda_n$ for $n\in\left\{1,2,\ldots,8\right\}$ (Figure~\ref{Fig:diagram}), with
\begin{equation}
  \lambda_0 = \sum_{n=1}^{8} \lambda_{n}.
  \label{Eq:total_jump_rate}
\end{equation}


We now make a few remarks. First, note that this description allows for jumps between diagonally neighbouring compartments. Second, the indexing of the jump rates starts at one for a jump to the right and continues in an anti-clockwise fashion, as shown in Figure~\ref{Fig:diagram}. Third, due to the reflective symmetries of a Cartesian mesh and since $D$ is spatially homogeneous, we have the following identities
\begin{equation}
	\lambda_1 = \lambda_5 \, ,
	\quad \lambda_3 = \lambda_7 \, ,
	\quad \lambda_2 = \lambda_4 = \lambda_6 = \lambda_8.
\end{equation}
Finally, note that in what follows we present derivations of the jump rates for a two-dimensional domain but that all derivations naturally extend to the three-dimensional case.



\subsubsection{The finite difference method}
\label{Sec:fdm}

The first method we consider to derive the diffusive jump rates is the FDM. Meinecke and L\"{o}tsedt~\cite{meinecke2016stochastic} chose a 9-point stencil which allows one to include a parameter $\alpha$ to study the effects of diagonal jumps, while maintaining a discretisation of the Laplacian that is still second order accurate~\cite{Strikwerda2004}. The standard 5-point stencil can be recovered by setting $\alpha = 0$ and hence neglecting diagonal jumps. We have
\begin{equation}
\Delta p \left( \textbf{x}_i, t \right) \approx
\delta^2_x \tilde{p}_i + \delta^2_y \tilde{p}_i
+\frac{1}{2} \alpha \kappa h^2 \delta^2_x \delta^2_y \tilde{p}_i,
\end{equation}
where $\alpha$ is a free parameter controlling the frequency of diagonal jumps, and the difference operators
are defined as follows:
\begin{eqnarray}
\nonumber
\delta^2_x \tilde{p}_i &=&	\frac{1}{\kappa^2 h^2} \left( \tilde{p}_{i+1} + \tilde{p}_{i-1} - 2 \tilde{p}_i \right); \\
\delta^2_y \tilde{p}_{i,j} &=&\frac{1}{h^2} \left( \tilde{p}_{i+n_x} + \tilde{p}_{i-n_x} - 2 \tilde{p}_i \right).
\end{eqnarray}
As such, the right-hand side of Equation~\eqref{Eq:diff_eq} can be approximated as
\begin{eqnarray}
\nonumber
	D \Delta p &\approx&\frac{D - D\alpha \kappa}{\kappa^2 h^2}
	\left( \tilde{p}_{i+1} + \tilde{p}_{i-1} \right)
	+ \frac{D \kappa - D \alpha}{\kappa h^2} \left(\tilde{p}_{i+n_x} +\tilde{p}_{i-n_x} \right)  \\
  \nonumber
	& & \quad + \frac{D\alpha}{2 \kappa h^2} \left(
  \tilde{p}_{i+n_x+1} +
  \tilde{p}_{i-n_x+1} +
  \tilde{p}_{i+n_x-1} +
  \tilde{p}_{i-n_x-1} \right) \\
	& & \quad - 2 D  \frac{\kappa^2 - \alpha \kappa + 1}{\kappa^2 h^2}  \tilde{p}_{i} \, .
\label{Eq:FDM_step}
\end{eqnarray}
Comparing the general expression for the discretised Laplacian in Equation~\eqref{Eq:discretised_laplacian} with the discretised Laplacian in Equation~\eqref{Eq:FDM_step} we have
\begin{eqnarray}
	\lambda_1 &=& \frac{D - D\alpha \kappa}{\kappa^2 h^2}, \\
	\lambda_2 &=& \frac{D\alpha}{2 \kappa h^2}, \\
	\lambda_3 &=& \frac{D \kappa - D \alpha}{\kappa h^2}, \\
	\lambda_0 &=& \frac{2 D (\kappa^2 - \alpha \kappa + 1)}{\kappa^2 h^2}.
\end{eqnarray}


\subsubsection{The finite volume method}
\label{Sec:fvm}

We now consider derivation of the jump rates using the FVM. We start by averaging the Laplacian at $\textbf{x}_{i}$
by integrating over compartment $i$
\begin{equation}
D \Delta p (\textbf{x}_{i}) \approx
\frac{D}{A} \int_{C_{i}} \Delta p \, \D \Omega,
\label{Eq:FVM_step_1}
\end{equation}
where $A$ is the area of the compartment $i$. Using the divergence theorem, we rewrite Equation~\eqref{Eq:FVM_step_1} as
\begin{equation}
D \Delta p (\textbf{x}_{i}) \approx \frac{D}{A}
\int_{\p C_i} \left(\textbf{n} \cdot \nabla \right) p \, \D s,
\label{Eq:FVM_step_2}
\end{equation}
where $\D s$ is a line element of the compartment boundary, $\p C_i$,
and $\textbf{n}$ is the normal vector to the compartment boundary at the point
$\textbf{x}$.

\begin{figure}
\centering
\includegraphics[scale=1.0]{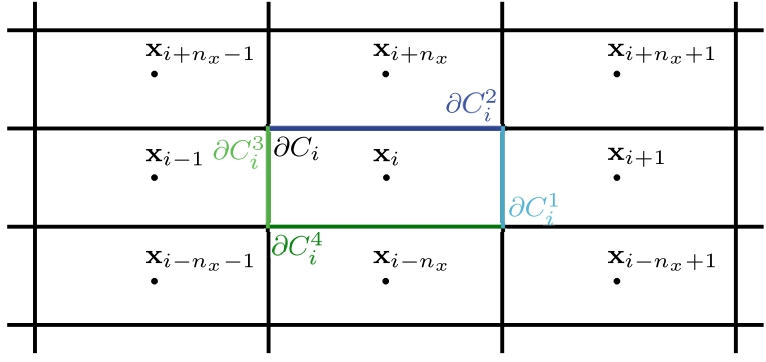}
\caption{Schematic diagram of how compartment $i$ boundary, $\p C_i$,  is split into four segments $\p C_i^k$ for $k = 1,2,3,4$.}
\label{Fig:fvm_diagram}
\end{figure}

Next, we split the integral in Equation~\eqref{Eq:FVM_step_2} into integrals along segments of compartment boundary $\p C_i$, as shown in Figure~\ref{Fig:fvm_diagram}:
\begin{equation}
  D \Delta p (\textbf{x}_{i}) \approx \frac{D}{A}
	\sum_{k=1}^4 \int_{\p C_i^k} \left(\textbf{n} \cdot \nabla \right) p \, \D s,
  \label{Eq:FVM_step_3}
\end{equation}
and we approximate the gradient of $p$ across the different parts of the boundary of the compartment using a finite difference method:
\begin{eqnarray}
\textbf{n} \cdot \nabla p \left\rvert_{\p C_i^1} \right. &\approx&
	\frac{\tilde{p}_{i+1} - \tilde{p}_{i}}{\kappa h};
  \label{Eq:FVM_step_4_1} \\
\textbf{n} \cdot \nabla p \left\rvert_{\p C_i^2} \right. &\approx&
  \frac{\tilde{p}_{i+n_x} - \tilde{p}_{i}}{h}  ;
  \label{Eq:FVM_step_4_2} \\
\textbf{n} \cdot \nabla p \left\rvert_{\p C_i^3} \right. &\approx&
	\frac{\tilde{p}_{i-1} - \tilde{p}_{i}}{\kappa h}  ;
  \label{Eq:FVM_step_4_3} \\
\textbf{n} \cdot \nabla p \left\rvert_{\p C_i^4} \right. &\approx&
	\frac{\tilde{p}_{i-n_x} - \tilde{p}_{i}}{h} .
  \label{Eq:FVM_step_4_4}
\end{eqnarray}
Substituting Equations~\eqref{Eq:FVM_step_4_1}--\eqref{Eq:FVM_step_4_4} into Equation~\eqref{Eq:FVM_step_3} gives
\begin{eqnarray}
	\lambda_1 &=& \frac{D}{\kappa^2 h^2} , \\
	\lambda_2 &=& 0 , \\
	\lambda_3 &=& \frac{D}{h^2} , \\
	\lambda_0 &=& \frac{2 D ( \kappa^2 + 1)}{\kappa^2 h^2} .
\end{eqnarray}


\subsubsection{The finite element method}
\label{Sec:fem}

The last numerical method we will use to derive the jump coefficients is the FEM. We multiply Equation~\eqref{Eq:diff_eq} by a test function $v(\textbf{x})$, which is a function in Sobolev space~\cite{Elman2010} (i.e. is bounded and has bounded first partial derivatives), and integrate over the whole domain $\Omega$:
\begin{eqnarray}
\nonumber
	\int_{\Omega} \frac{\p p}{\p t} v \, \D x \, \D y
	&=& D \int_{\Omega} v \Delta p \, \D x \, \D y  \\
	&=& D \int_{\Omega} \left[ \nabla \cdot \left( v \nabla p \right) -
	\nabla p \cdot \nabla v \right] \, \D x \, \D y.
	\label{Eq:FEM_2d_step_1}
\end{eqnarray}
Now we use the divergence theorem to give
\begin{equation}
	\int_{\Omega} \frac{\p p}{\p t} v \,\D x \,\D y =
	D \int_{\p \Omega} \textbf{n} \cdot \left[ v \nabla p \right] \D s -
	D \int_{\Omega} \nabla p \cdot \nabla v \,\D x \, \D y,
	\label{Eq:FEM_2d_step_2}
\end{equation}
where $\D s$ is the line element of the domain boundary, $\p \Omega$, and $\textbf{n}$ is the normal vector to the domain boundary at the point $\textbf{x}$. Due to the Neumann boundary conditions, the first term in Equation~\eqref{Eq:FEM_2d_step_2} vanishes and we are left with
\begin{equation}
	\int_{\Omega} \frac{\p p}{\p t} v \, \D x \, \D y =
	- D \int_{\Omega} \nabla p \cdot \nabla v \, \D x \, \D y.
	\label{Eq:FEM_2d_step 3}
\end{equation}
We express the functions $p$, $\p p / \p t$ and $v$ in terms of the basis functions $\varphi_j$,
\begin{eqnarray}
	v(x, y) &=& \varphi_j(x, y), \vphantom{\sum_{i=1}^I}
	\label{Eq:FEM_2d_test_function} \\
	p(x, y, t) &\approx& \sum_{i=1}^I \tilde{p}_i(t) \varphi_i(x, y) ,
	\label{Eq:FEM_2d_u} \\
	\frac{\p p}{\p t} &\approx& \sum_{i=1}^I \frac{\D  \tilde{p}_i}{\D  t} (t) \varphi_i(x, y)
	\label{Eq:FEM_2d_dudt} ,
\end{eqnarray}
where the $\varphi_i$ are defined as
\begin{equation}
	\varphi_i \left( x, y \right) =
	\left\{
	\begin{array}{cc}
		\left(1-\cfrac{x-x_i}{\kappa h}\right) \left(1-\cfrac{y-y_i}{h}\right) &
		 \text{for} \quad \textbf{x} \in
		\left[ x_i , x_i + \kappa h \right] \times \left[ y_i , y_i + h \right], \\
		\left(1+\cfrac{x-x_i}{\kappa h}\right) \left(1-\cfrac{y-y_i}{h}\right) &
		 \text{for} \quad \textbf{x} \in
		\left[ x_i - \kappa h , x_i \right] \times \left[ y_i , y_i + h \right], \\
		\left(1+\cfrac{x-x_i}{\kappa h}\right) \left(1+\cfrac{y-y_i}{h}\right) &
		 \text{for} \quad \textbf{x} \in
		\left[ x_i - \kappa h , x_i \right] \times \left[ y_i - h , y_i \right], \\
		\left(1-\cfrac{x-x_i}{\kappa h}\right) \left(1+\cfrac{y-y_i}{h}\right) &
		 \text{for} \quad \textbf{x} \in
		\left[ x_i, x_i + \kappa h \right] \times \left[ y_i - h , y_i \right], \\
		0 & \text{otherwise},
	\end{array}
	\right.
\end{equation}
for $i\in \left\lbrace 1, \ldots, I \right\rbrace$, and $\varphi_j ( \textbf{x} ) = 0$ for $\textbf{x} \notin \Omega$. We choose linear basis functions for simplicity, however, any form will suffice, as long as the following condition is met:
\begin{equation}
	\varphi_j \left( x_i, y_i \right) =
	\begin{cases}
		1 \quad \text{if} \quad i=j  ;\\
		0 \quad \text{if} \quad i \neq j  .
	\end{cases}
\end{equation}
Substituting Equations~\eqref{Eq:FEM_2d_test_function},~\eqref{Eq:FEM_2d_u}
and~\eqref{Eq:FEM_2d_dudt} into Equation~\eqref{Eq:FEM_2d_step 3} gives, for each $j$,
\begin{equation}
	\sum_{i=1}^I \left( \int_{\Omega} \varphi_i \varphi_j \,
	\D x \, \D y \right) \frac{\D  \tilde{p}_i}{\D  t} =
	- \sum_{i=1}^I \left( \int_{\Omega} D \nabla \varphi_i \cdot
	\nabla \varphi_j \,\D x \,\D y \right) \tilde{p}_i ,
\end{equation}
which can be expressed as a matrix equation
\begin{equation}
	M \frac{\D  \tilde{\textbf{p}}}{\D t} = - N \tilde{\textbf{p}} \, ,
	\label{Eq:FEM_matrix_equation}
\end{equation}
where
\begin{eqnarray}
\tilde{\textbf{p}} &=& (\tilde{p}_1, \tilde{p}_2, \ldots, \tilde{p}_I), \vphantom{\int_{\Omega}} \\
M_{ij} &=& \int_{\Omega} \varphi_i \, \varphi_j \,\D x\,\D y, \\
N_{ij} &=& \int_{\Omega} D \nabla \varphi_i \cdot \nabla \varphi_j \,\D x\,\D y.
\end{eqnarray}
Evaluating the elements of the matrices $M$ and $N$, and using the lumped mass matrix approach (see Supplementary Information), we finally arrive at the following expressions for the jump coefficients:
\begin{eqnarray}
	\lambda_1 &=& \frac{D (2 - \kappa^2)}{3 \kappa^2 h^2}; \\
	\lambda_2 &=& \frac{D ( \kappa^2 + 1 )}{6 \kappa^2 h^2};\\
	\lambda_3 &=& \frac{D ( 2 \kappa^2 - 1 ) }{3 \kappa^2 h^2};\\
	\lambda_0 &=& \frac{4 D (\kappa^2 + 1)}{3\kappa^2 h^2}.
\end{eqnarray}


\begin{table}[t]
\centering
\begin{TAB}(r,0cm,0.1cm)[0.5cm]{|c|c|c|c|}{|c|c|c|c|}
Jump rates & FDM & FVM & FEM \\
$\lambda_1$ &
$ \displaystyle \frac{D - D\alpha \kappa}{\kappa^2 h^2}$ &
$ \displaystyle \frac{D}{\kappa^2 h^2}$ &
$ \displaystyle \frac{D (2 - \kappa^2)}{3 \kappa^2 h^2}$ \\
$\lambda_2$ &
$ \displaystyle \frac{D\alpha}{2 \kappa h^2}$ &
$ \displaystyle 0$ &
$ \displaystyle \frac{D ( \kappa^2 + 1 )}{6 \kappa^2 h^2}$ \\
$\lambda_3$ &
$ \displaystyle \frac{D \kappa - D \alpha}{\kappa h^2}$ &
$ \displaystyle \frac{D}{h^2}$ &
$ \displaystyle \frac{D ( 2 \kappa^2 - 1 ) }{3 \kappa^2 h^2}$
\end{TAB}
\caption{Comparison of jump rates from different derivations.}
\label{Table_jump_ceofficients}
\end{table}


\subsubsection{The first exit time method}
\label{Sec:fet}

\begin{figure}
\centering
\includegraphics[scale=1.0]{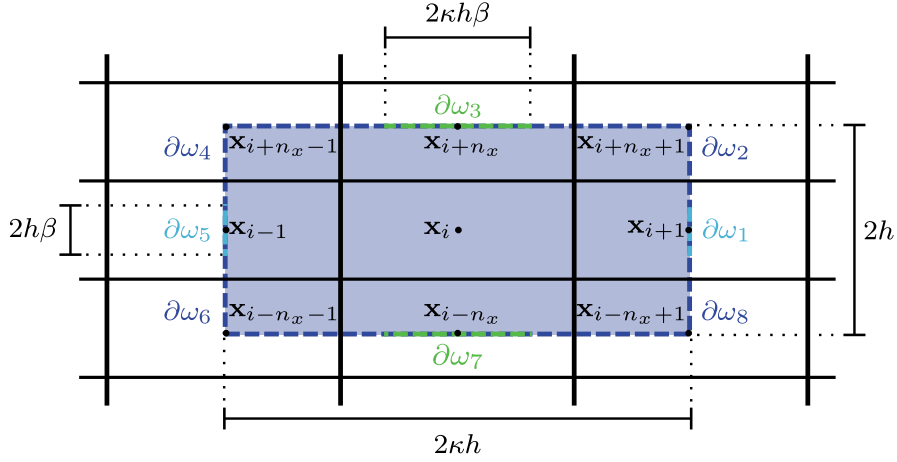}
\caption{Diagram of the domain $\omega_{i}$, which has been shaded in dark blue. The boundary of the domain $\omega_{i}$, indicated by the dashed line, is split into eight sections coloured as follows:
$\p\omega_1$ is the rightmost light blue;
$\p\omega_2$ is the upper-right dark blue;
$\p\omega_3$ is the uppermost green;
$\p\omega_4$ is the upper-left dark blue;
$\p\omega_5$ is the leftmost light blue;
$\p\omega_6$ is the lower-left dark blue;
$\p\omega_7$ is the lower green; and
$\p\omega_8$ is the lower-right dark blue.}
\label{Fig:fet_diagram}
\end{figure}

Meinecke and L\"ostedt proposed an additional method of calculating the jump rates~\cite{Lotstedt2015, meinecke2016stochastic}. This method involves considering the FET distribution from a domain $\omega_i$ (centered on  $C_i$) of a single molecule undergoing Brownian motion that is placed initially at the centre of $C_i$. Note that to model diffusion in a RDME framework using FET-derived jump rates we approximate the probability distribution function (PDF) for the FET by an exponential distribution with the same mean. We follow the choice of domain $\omega_{i}$ from~\cite{meinecke2016stochastic}, where $\omega_{i}$ is
defined by its boundary, $\p \omega_{i}$, and taken to be a rectangle that intersects the centres of neighbouring compartments of compartment $i$, as shown in Figure~\ref{Fig:fet_diagram}, so that $\omega_{i}$ completely contains the compartment $i$\footnote{The choice of $\omega_{i}$ has been briefly discussed in \cite{Lotstedt2015} and so we do not explore it further here.}.


We follow Meinecke and L\"ostedt~\cite{meinecke2016stochastic} in choosing jump rates of the form
\begin{equation}
\lambda_n=\lambda_0\,\theta_n,
\label{Eq:FET_jump_coefficients}
\end{equation}
where $\theta_n$ is the probability that a molecule exits $\omega_i$ into the $n$th neighbouring compartment (see Figure~\ref{Fig:diagram}). We define a molecule to exit $\omega_i$ into the $n$th neighbouring compartment if it exits through the segment of the domain boundary denoted by $\p\omega_i^n$ in Figure~\ref{Fig:fet_diagram}. There is flexibility in the choice of discretisation of the boundary $\p \omega_{i}$ into segments $\p\omega_i^n$, $n=1,\ldots,8$, and we define the discretisation using parameter $\beta$ as follows:
\begin{equation}
\begin{split}
	&\left| \p \omega_1 \right| = \left| \p \omega_5 \right| = 2 \beta h;\\
	&\left| \p \omega_3 \right| = \left| \p \omega_1 \right| = 2 \beta \kappa h; \\
  &\left| \p \omega_2 \right| = \left| \p \omega_4 \right| =
  \left| \p \omega_6 \right| = \left| \p \omega_8 \right| = (1 - \beta )(1 + \kappa)h.
\end{split}
\end{equation}
We will explore the effect of changing the parameter $\beta$ on the jump rates in Section~\ref{Sec:comparison}.

We proceed by considering two random variables associated with a molecule governed by Brownian motion exiting from the domain $\omega_{i}$: $\tau$ and $\eta$. The random variable $\tau$ denotes the FET out of $\omega_i$, whereas $\eta$ denotes the exit segment. We use $\tau$ to calculate $\lambda_0$ by, as explained earlier, equating the mean of the PDF of $\tau$ with the rate parameter of the exponential distribution used in Algorithm~\ref{Alg:SSA}. The random variable $\eta$, on the other hand, is used to calculate $\theta_n$.

We calculate the PDF of $\tau$ using the FPE (as outlined in Section~\ref{Sec:diffusion}) that describes the probability for the molecule to be at position $\textbf{x}$ at time $t$:
\begin{equation}
\frac{\p p}{\p t} =
D \Delta p \quad \text{for} \quad \textbf{x} \in \omega_{i},
\label{Eq:fet_main_equation}
\end{equation}
where $D$ is the diffusion coefficient. To calculate the PDF of the FET, we use Dirichlet (absorbing) boundary conditions
\begin{equation}
	p(\textbf{x}, t) =
	0 \quad \text{for} \quad \textbf{x} \in \p \omega_{i},
  \label{Eq:fet_bc}
\end{equation}
and assume that initially the molecule at the centre of $\omega_{i}$ so that
\begin{equation}
	p(\textbf{x}, 0) =
	\delta (\textbf{x} - \textbf{x}_{i}).
	\label{Eq:fet_ic}
\end{equation}

Using $p(\textbf{x}, t)$ we calculate the survival probability, $S(t)$, which is the probability that a molecule remains in the domain $\omega_{i}$ at least until time $t$, as~\cite{Redner2002}
\begin{equation}
	S(t) = \mathbb{P}( \tau \geqslant t) =
	\int_{\omega_{i}} p(\textbf{x}, t) \, \D \omega,
	\label{Eq:survival_probability}
\end{equation}
where $\D \omega$ is a domain element of the domain $\omega_{i}$. Now, given $S(t)$, we derive the probability
that a molecule exits $\omega_{i}$ at time $t$, i.e. the PDF of the random variable $\tau$, denoted by $\mathbb{P}(\tau = t)$, as
\begin{equation}
	\mathbb{P}(\tau = t) = - \frac{\p S}{\p t} \, .
  \label{Eq:pdf_tau}
\end{equation}
Substituting Equation~\eqref{Eq:survival_probability} into Equation~\eqref{Eq:pdf_tau}, results in
\begin{eqnarray}
	\mathbb{P}(\tau = t)
	\nonumber
		&=& - \int_{\omega_{i}} \frac{\p p}{\p t} \,\D \omega  \\
		\nonumber
		&=& -D \int_{\omega_{i}} \Delta p \,\D \omega  \\
		&=& -D \int_{\p \omega_{i}} \textbf{n} \cdot \nabla p \,\D s  ,
\label{Eq:pdf_fot_tau}
\end{eqnarray}
where $\D s$ is a line element of $\p \omega_{i}$. Finally, we calculate the mean of the PDF for $\tau$ as
\begin{equation}
	E_\tau = \mathbb{E} \left[ \tau \right] =
	\int_0^\infty t \,\mathbb{P}(\tau = t) \,\D t =
	\int_0^\infty S(t) \,\D t ,
  \label{Eq:FET_lambda_0_expression_1}
\end{equation}
and set
\begin{equation}
 \lambda_0  = \frac{1}{E_\tau}.
  \label{Eq:fet_means_identity}
\end{equation}

To calculate the PDF for $\eta$, we use the law of total probability
\begin{equation}
  \mathbb{P}(\eta = n) =
  \int_0^\infty \mathbb{P}(\eta = n | \tau = t)\mathbb{P}(\tau = t) \,\D t .
  \label{Eq:law_of_total_probability}
\end{equation}
Following \cite{meinecke2016stochastic}, we have the probability of a molecule exiting through segment $\p\omega_i^n$ given that a molecule exits at time $t$, i.e. the probability $\mathbb{P}(\eta = n| \tau = t)$,
as follows
\begin{equation}
  \mathbb{P}(\eta = n | \tau = t ) =
  \frac{\int_{\p\omega_i^n} \textbf{n} \cdot \nabla p \,\D s}{\int_{\p\omega_i} \textbf{n} \cdot \nabla p \,\D s}.
  \label{Eq:conditional_probability}
\end{equation}
Therefore, substituting Equation~\eqref{Eq:pdf_fot_tau} and Equation~\eqref{Eq:conditional_probability} into Equation~\eqref{Eq:law_of_total_probability},
we have
\begin{eqnarray}
  \nonumber
  \theta_n &=& \mathbb{P}(\eta = n) \vphantom{\int_0^\infty} \\
  \nonumber
  &=& \int_0^\infty\frac{\int_{\p\omega_i^n}\textbf{n} \cdot \nabla p \,\D s}{\int_{\p\omega_i} \textbf{n} \cdot \nabla p \D s} \left(-D \int_{\p \omega_{i}} \textbf{n} \cdot \nabla p \,\D s\right) \D t \\
  &=& -D \int_0^\infty \int_{\p\omega_i^n} \textbf{n} \cdot \nabla p \,\D s \,\D t .
\label{Eq:theta_def}
\end{eqnarray}
Therefore, given solutions to Equation~\eqref{Eq:FET_jump_coefficients} and
Equation~\eqref{Eq:fet_main_equation}, we can calculate the jump rates $\lambda_n$ for $n=1,\ldots,8$. To see the full expressions for the jump coefficients derived using the FET method see the Supplementary Information.


\subsubsection{Comparison of the diffusive jump rates}
\label{Sec:comparison}


\begin{figure}
\centering
\includegraphics[width=\textwidth]{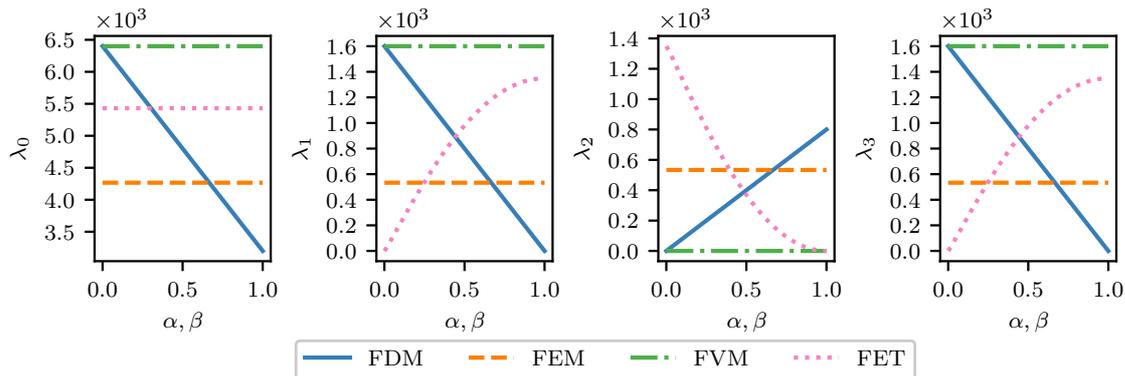}
\caption{Plot of the jump rates as a function of the implementation parameters, $\alpha$ and $\beta$, for $h = 0.025 \, \mu m$, $\kappa = 1.0$ and $D = 1.0 \, \mu m^2 s^{-1}$ for the total jump rate, $\lambda_0$, as well as $\lambda_1$, $\lambda_2$ and $\lambda_3$. Since the derivations using the FEM and the FVM do not have any implementation parameters, the jump rates are constant for the given simulation parameters.}
\label{Fig:All_lambdas}
\end{figure}


Figure~\ref{Fig:All_lambdas} shows how the four methods of derivation compare to one another as the parameters $\alpha$ and $\beta$, introduced during the derivations of the FDM and FET jump rates, respectively, are varied. Both $\alpha$ and $\beta$ control the frequency of diagonal jumps. For the FDM-derived jump rates, increasing the parameter $\alpha$ increases $\lambda_2$, the rate of diagonal jumps. On the other hand, increasing $\beta$ has the opposite effect on $\lambda_2$ in the FET case. As the FEM- and the FVM-derived jump rates do not have any free parameters, they are constant in each of the plots.


\subsection{Derivation of the diffusive jump rates on a growing domain}

The SDE description of a molecule undergoing Brownian motion on a two-dimensional growing domain $\Omega_t=[0, L(t)]\times[0, L(t)]$ we use in this work is
\begin{eqnarray}
x(t+\D t) &=& x(t) + v_x(x, y, t) \D t + \sqrt{2D} \D{W}_x , \\
y(t+\D t) &=& y(t) + v_y(x, y, t) \D t + \sqrt{2D} \D{W}_y ,
\end{eqnarray}
where the drift terms, $v_x(x, y, t)$ and $v_y(x, y, t)$, describe the effects of domain growth. The corresponding FPE is
\begin{equation}
  \frac{\p p}{\p t} = D \nabla^2 p - \nabla \cdot (\textbf{v} p)
  \quad \text{for} \quad \textbf{x} \in \Omega_t,
  \label{Eq:main_pde_2d}
\end{equation}
with Neumann boundary conditions
\begin{equation}
  \left. \nabla p \right|_{\p \Omega_t} = 0,
\end{equation}
and initial condition
\begin{equation}
  p(\textbf{x}, 0) = p_0(\textbf{x}) \quad \text{for} \quad \textbf{x} \in \Omega_0.
\end{equation}
Assuming growth is isotropic, the drift terms, $\textbf{v}(\textbf{x}, t)=(v_x(x, y, t),v_y(x, y, t))$, are given by
\begin{equation}
  \textbf{v}(\textbf{x}, t) =
  \left( \frac{x \dot{L}(t)}{L(t)}, \frac{y \dot{L}(t)}{L(t)} \right).
  \label{Eq:v_2d}
\end{equation}

Care now must be taken when considering the equation that is to be used to calculate the diffusive jump rates, since on the growing domain the compartments are also changing in size, with $x$ and $y$ dimensions $\kappa{h(t)}=L(t)/n_x$ and $h(t)=L(t)/n_y$, respectively. This means that the probability to find the particle in compartment $C_i(t)$ is given by $\int_{C_i(t)}p(\textbf{x},t)\D\textbf{x}$ with
\begin{eqnarray}
\frac{\D}{\D{t}}\int_{C_i(t)}p(\textbf{x},t)\,\D\textbf{x}
&=&
\int_{C_i(t)}\frac{\p }{\p t}p(\textbf{x},t)\,\D\textbf{x}
+\int_{\p C_i(t)}(\textbf{v}\cdot\textbf{n})\,p(\textbf{x},t)\,\D\textbf{x}.
\end{eqnarray}
Using equation~\eqref{Eq:main_pde_2d} together with the divergence theorem, we see that the terms involving the drift, $\textbf{v}$, cancel so that we should consider the particle only as diffusing on the growing domain. We employ a change of coordinate to transform to a static domain,
\begin{equation}
  (x, y, t) \rightarrow (\xi, \eta, \tau) =
  \left( \frac{x}{L(t)}, \frac{y}{L(t)}, t \right).
  \label{Eq:transformed_coord_2d}
\end{equation}
so that we can derive the diffusive jump rates using the diffusion equation with time-varying diffusion coefficient:
\begin{equation}
\frac{\p p}{\p \tau} = \frac{D}{L(t)^2} \nabla_\xi \, p
\quad \text{for} \quad \bm{\xi} \in \Omega_0,
  \label{Eq:transformed_main_pde_2d}
\end{equation}
where $\bm{\xi} = (\xi, \eta)$ and $\nabla_\xi = (\p / \p \xi, \p / \p \eta)$.

The key advantage is that we can now simply use the diffusive jump rate derivations for the static case for the FDM, FVM and FEM (see Table~\ref{Table_jump_coefficients_growing}). Note that we no longer consider the FET approach to deriving the diffusive jump rates. Simpson \textit{et al.}~\cite{simpson2015exactcalc} have previously shown that on a growing domain the survival probability does not approach zero in the limit $t\to\infty$. As a result, the FET distribution has infinite mean so we cannot use it to parameterise the exponential distribution as required in Equation~\eqref{Eq:fet_means_identity}.


\begin{table}[t]
\centering
\begin{TAB}(r,0cm,0.1cm)[0.5cm]{|c|c|c|c|}{|c|c|c|c|}
Jump rates & FDM & FVM & FEM \\
$\lambda_1(t)$ &
$ \displaystyle \frac{D - D\alpha \kappa}{\kappa^2 h^2(t)}$ &
$ \displaystyle \frac{D}{\kappa^2 h^2(t)}$ &
$ \displaystyle \frac{D (2 - \kappa^2)}{3 \kappa^2 h^2(t)}$ \\
$\lambda_2(t)$ &
$ \displaystyle \frac{D\alpha}{2 \kappa h^2(t)}$ &
$ \displaystyle 0$ &
$ \displaystyle \frac{D ( \kappa^2 + 1 )}{6 \kappa^2 h^2(t)}$ \\
$\lambda_3(t)$ &
$ \displaystyle \frac{D \kappa - D \alpha}{\kappa h^2(t)}$ &
$ \displaystyle \frac{D}{h^2(t)}$ &
$ \displaystyle \frac{D ( 2 \kappa^2 - 1 ) }{3 \kappa^2 h^2(t)}$
\end{TAB}
\caption{Comparison of jump rates from different derivations on a growing domain.}
\label{Table_jump_coefficients_growing}
\end{table}


\section{Results}
\label{Sec:results}

The SSA outlined in Algorithm~\ref{Alg:SSA} of Section~\ref{Sec:model} allows us to simulate a range of reaction-diffusion systems and test the impact of the different diffusive jump derivations on model predictions. We first explore a range of systems on a static domain, before moving to a growing domain.


\subsection{Static domain}
\label{Sec:static_domain}

For simple systems, e.g. those that contain only zeroth and first order reactions, it is possible to compare results with analytic solutions of the corresponding macroscopic partial differential equation (PDE) models (see Supplementary Information for more details of the models and their analytic solution). The PDE represents the evolution of molecule concentration in time, so to compare with the results of stochastic simulation, we use the following error measure:
\begin{equation}
e(t)=\sqrt{\sum_{i=1}^{I}A(t)\left|\frac{U_i(t)}{A(t)}-u(\textbf{x}_{i},t)\right|^2},
\label{Eq:error}
\end{equation}
where $A(t)$ is the compartment area at time $t$, and $u(\textbf{x}_i,t)$ is the analytical solution of the corresponding macroscale PDE at $\textbf{x}_i$ at time $t$.


\begin{figure}
\centering
\includegraphics[width=\textwidth]{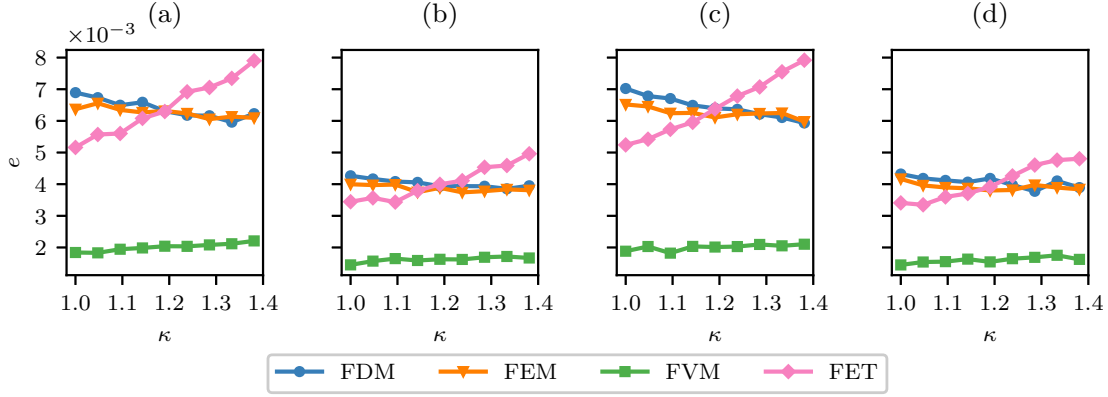}
\caption{Plot of the error, as defined in Equation~\eqref{Eq:error}, as a function of $\kappa$, the compartment aspect ratio, for simulations with (a) diffusion only,	(b) diffusion and production, (c) diffusion and decay and (d) diffusion, decay and production.	The domain $\Omega = [0, 20]\times[0, 20]$ is discretised into $21 \times 21\kappa$ compartments and $5\times10^6$ molecules are placed initially in the central compartment. Parameters are: $D=1 \, \mu m^2 s^{-1}$, $\alpha = 0.7$, $\beta = 0.5$, $T=5 \, s$ and $k_1 = 100 \, \mu m^{-2} s^{-1}$, $k_{-1} = 0.1 \, s^{-1}$ for the systems involving production and decay, respectively.}
\label{Fig:plot_of_error_against_kappa}
\end{figure}


\subsubsection{Diffusion}
\label{Sec:results_diffusion}

We first investigate the effects of varying the compartment aspect ratio, $\kappa$. The macroscale PDE that describes the evolution of molecule concentration is
\begin{equation}
\frac{\p u(\textbf{x}, t)}{\p t} = D \Delta u (\textbf{x}, t)
\quad \text{for} \quad \textbf{x} \in \Omega,
\label{Eq:diff_eq_results}
\end{equation}
with Neumann boundary conditions
\begin{equation}
\left( \textbf{n} \cdot \nabla \right) p = 0 \quad \text{for} \quad \textbf{x} \in \partial \Omega,
\end{equation}
where $\textbf{n}$ is the normal vector to the domain boundary $\p \Omega$. We choose the initial condition to be
\begin{equation}
    u(\textbf{x}, 0) =
    \left\{
    \begin{array}{cc}
        5\times10^6 / A & \text{for } 0 < x < \kappa h, 0 < y < h, \\
        0 & \text{otherwise}.
    \end{array}
    \right.
\end{equation}
We use the analytic solution of this PDE (see Supplementary Information) to calculate the error, $e$, using Equation~\eqref{Eq:error}. Figure~\ref{Fig:plot_of_error_against_kappa}(a) shows that the error is relatively unaffected by variation in the compartment aspect ratio, $\kappa$. For the values of $\alpha$ and $\beta$ used, the FVM-derived jump rates give rise to the smallest error and all other methods (FDM, FEM and FET) give rise to errors that are very similar to each other. Note that for $\alpha=0$ the diffusive jump rates for the FDM and the FVM are identical, indicating that increasing $\alpha$ increases the error.


\subsubsection{Zeroth and first order reactions}
\label{Sec:results_first_order}

We now consider reaction-diffusion systems involving the decay of molecules,
given by the chemical equation
\begin{equation}
\text{U} \xrightarrow{k_{-1}} \emptyset,
\end{equation}
and the production of molecules, given by the chemical equation
\begin{equation}
\emptyset \xrightarrow{k_1} \text{U},
\end{equation}
alongside diffusion. The corresponding propensity functions take the form
\begin{eqnarray}
a_i^{\text{decay}} (\textbf{u}(t)) &=& k_{-1} u_{i}(t) , \\
a_i^{\text{prod}} (\textbf{u}(t)) &=& k_{1} A(t),
\end{eqnarray}
for the decay reaction and the production reaction in compartment $i$, respectively. As before, $A(t)$ is the area of a compartment at time $t$, and $u_i(t)$ is the number of molecules in compartment $i$ at time $t$.
We consider production-only, decay-only and production-and-decay systems, and the macroscale PDEs that describe the evolution of molecule concentration are, respectively,
\begin{eqnarray}
\frac{\p u(\textbf{x}, t)}{\p t} &=& D \Delta u (\textbf{x}, t) + k_1, \label{Eq:prod} \\
\frac{\p u(\textbf{x}, t)}{\p t} &=& D \Delta u (\textbf{x}, t) - k_{-1} u, \label{Eq:decay} \\
\frac{\p u(\textbf{x}, t)}{\p t} &=& D \Delta u (\textbf{x}, t) + k_1 - k_{-1} u. \label{Eq:prod_decay}
\end{eqnarray}
Again, we use Neumann boundary conditions
\begin{equation}
\left( \textbf{n} \cdot \nabla \right) p = 0 \quad \text{for} \quad \textbf{x} \in \partial \Omega,
\end{equation}
where $\textbf{n}$ is the normal vector to the domain boundary $\p \Omega$, and initially $5\times10^6$ molecules are placed in the central compartment.

Figures~\ref{Fig:plot_of_error_against_kappa}(b)--(d) show how the error, $e$, varies with $\kappa$ for these three systems (solutions to the PDEs in Equations~\eqref{Eq:prod}--\eqref{Eq:prod_decay} are given in the Supplementary Information). Again, the FVM out-performs the other three methods (FDM, FEM and FET) for the chosen values of $\alpha$ and $\beta$. In addition, for systems involving decay (a first order reaction) the error increases as the aspect ratio of the compartments, $\kappa$, increases.


\subsubsection{Second-order reactions}
\label{Sec:results_2nd_order}

\begin{figure}
\centering
\includegraphics{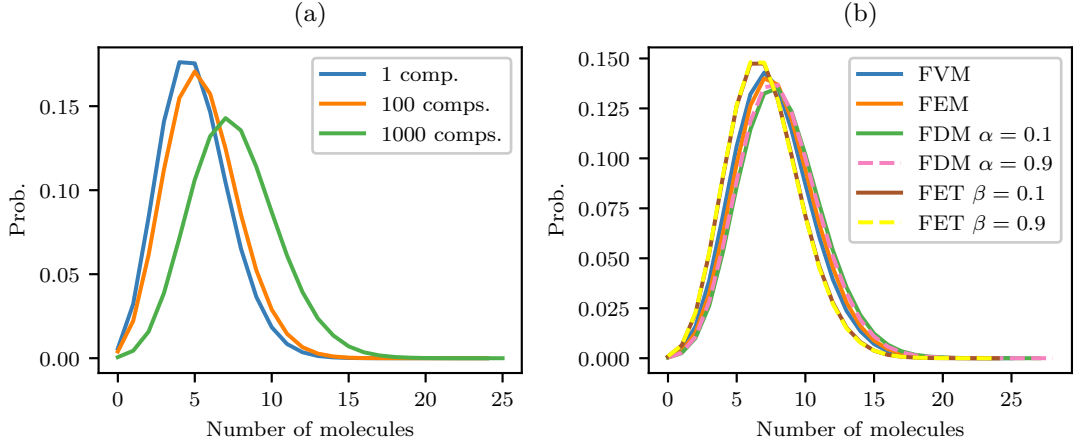}
\caption{Stationary distributions as calculated from simulations run until $T=1\times10^6 s$, on the domain $\Omega = [0,1]\times[0,1]$, with the following reaction rates: $k_1 = 0.2 \, \mu m^2 s^{-1}$ and $k_2 = 1.0 \, \mu m^{-2} s^{-1}$. We initialised the simulations with five molecules of U and one molecule of V distributed at random across the compartments. (a) Comparison of the stationary distributions as the compartment number is varied, where we used FVM-derived jump rates. (b) Comparison of the stationary distributions for different methods of derivation of the diffusive jump rates. The domain has been discretised into $1000\times1000$ compartments.}
\label{Fig:stationary_dists}
\end{figure}

To investigate the effects of different derivations of the jump rates in systems involving second-order reactions, we now consider a system composed of two species, U and V, which undergo diffusion (with rates $D_u$ and $D_v$, respectively) and the following reactions
\begin{equation}
\text{U} + \text{V} \xrightarrow{k_1} \text{V}, \quad \emptyset \xrightarrow{k_2} \text{U},
\label{Eq:two_species_decay_system}
\end{equation}
\textit{i.e.} U is consumed upon contact with V at rate $k_1$, and U is produced at rate $k_2$. The corresponding propensity functions take the form
\begin{eqnarray}
a_i^{1} (\textbf{u}(t)) &=& \frac{k_1}{A}u_i(t)v_i(t),\\
a_i^{2} (\textbf{u}(t)) &=& k_2A(t).
\end{eqnarray}
We cannot write down equivalent closed-form macroscale PDEs for the evolution of U and V concentration over time because the system contains second order reactions. As such, we investigate how the stationary distribution, estimated from stochastic simulations, is affected by the choice of compartment size, for the four different methods of deriving the diffusive jump rates. In Figure~\ref{Fig:stationary_dists}(a) we see that as the number of compartments increases the stationary distribution shifts to the right, towards higher molecule numbers. This is because the rate of bimolecular reactions decreases as the number of compartments increases~\cite{Isaacson2013}. Figure~\ref{Fig:stationary_dists}(b) shows how the stationary distribution changes with the method of derivation of the diffusive jump rates. All methods give slightly different results; in particular, using $\beta = 0.1$ or $\beta = 0.9$ in the FEM results in stationary distributions that are shifted noticeably to the left.


\subsubsection{Schnakenberg kinetics}
\label{Sec:results_schnakenberg_kinetics}

Finally, we investigate the effects of different derivations of diffusive jump rates on Turing's reaction-diffusion model of pattern formation~\cite{Turing1952}. We focus on Schnakenberg kinetics~\cite{gierer1972theory, schnakenberg1979simple} as they are simple to simulate. The system is composed of two species, U and V, which undergo diffusion (with rates $D_u$ and $D_v$, respectively) and the following reactions
\begin{equation}
\emptyset \xrightarrow{k_1} \text{U}, \quad
\text{U} \xrightarrow{k_2} \emptyset, \quad
2\text{U} + \text{V} \xrightarrow{k_3} 3\text{U}, \quad
\emptyset \xrightarrow{k_4} \text{V}.
\label{Eq:schnakenberg_system}
\end{equation}
The corresponding propensity functions take the form
\begin{eqnarray}
a_i^{1} (\textbf{u}(t),\textbf{v}(t)) &=& k_1A(t),\\
a_i^{2} (\textbf{u}(t),\textbf{v}(t)) &=& k_2u_i(t),\\
a_i^{3} (\textbf{u}(t),\textbf{v}(t)) &=& \frac{k_3}{A(t)^2}u_i(t)(u_i(t)-1)v_i(t),\\
a_i^{4} (\textbf{u}(t),\textbf{v}(t)) &=& k_4A(t).
\end{eqnarray}

We quantify the patterns formed using the discrete Fourier cosine transform. Let $f(x, y) : \mathbb{R}^2 \rightarrow \mathbb{R}$, then the discrete Fourier cosine transform is defined as
\begin{equation}
\hat{f} (k_x, k_y) =
\Delta_x \Delta_y \sum_{i=1}^{n_x} \sum_{j=1}^{n_y}
\cos ( k_x \Delta_x (i-1) ) \cos ( k_y \Delta_y (j-1) ) f (x, y) ,
\label{Eq:cosine_transform}
\end{equation}
where $\Delta_x$ is the spacing between points $\textbf{x}_i$ and $\textbf{x}_{i+1}$ and $\Delta_y$ is the spacing between points $\textbf{x}_i$ and $\textbf{x}_{i+n_x}$ (here,  $\Delta_x = \kappa h$ and $\Delta_y = h$). The
wavenumbers, $k_x$ and $k_y$, give the spatial frequency of a pattern, with wavemodes $m_x$ and $m_y$ such that $k_x = m_x \pi / L$ and $k_y = m_y \pi / L$, where $L$ is the side length of the square domain $\Omega$~\cite{woolley2011stochastic}. To showcase results, we plot the power spectrum, defined as
\begin{equation}
P_s	= \left| \widehat{f}(k_x, k_y) \right|^2,
\end{equation}
where $\widehat{f}(\cdot, \cdot)$ is the discrete Fourier cosine transform defined in Equation~\eqref{Eq:cosine_transform}.

We can predict which patterns are possible in equivalent PDE models of pattern formation using linear stability analysis (LSA)~\cite{murray2001mathematical} (see the Supplementary Information for more details). Woolley \textit{et al.}~\cite{woolley2011power} developed a method using a weak noise expansion (WNE) of the RDME~\cite{vanKampen2007} to determine the wavemodes that can evolve in stochastic simulations (see the Supplementary Information for more details). The advantage of the WNE approach over LSA in determining whether a particular wavemode is possible for a given parameter set is that the WNE takes into account the details of the diffusive jump rates. We compare the simulation results to the predictions of the WNE and LSA
to see whether the patterns formed fall within the ranges of wavemodes predicted by those methods.

All the simulations are run until $T = 1800 \, s$ on the domain $\Omega = [0, 1] \times [0, 1]$, which is discretised into $40 \times \lfloor 40\kappa \rfloor$ compartments. We use the following reaction rates:
$D_u = 1\times10^{-5}  \, \mu m^2 s^{-1}$, $D_v = 1\times10^{-3} \, \mu m^2 s^{-1}$, $k_1= 1 \, s^{-1}$, $k_2 = 0.02 \, s^{-1}$, $k_3 = 1 \times 10^{-6} \, s^{-1}$ and $k_4= 3 \, s^{-1}$. We initialise the system at the spatially uniform steady state, placing 200 molecules of species U and 75 molecules of species V in each compartment.



\pagebreak

\paragraph{The finite difference method}

\begin{figure}[h]
\centering
\includegraphics{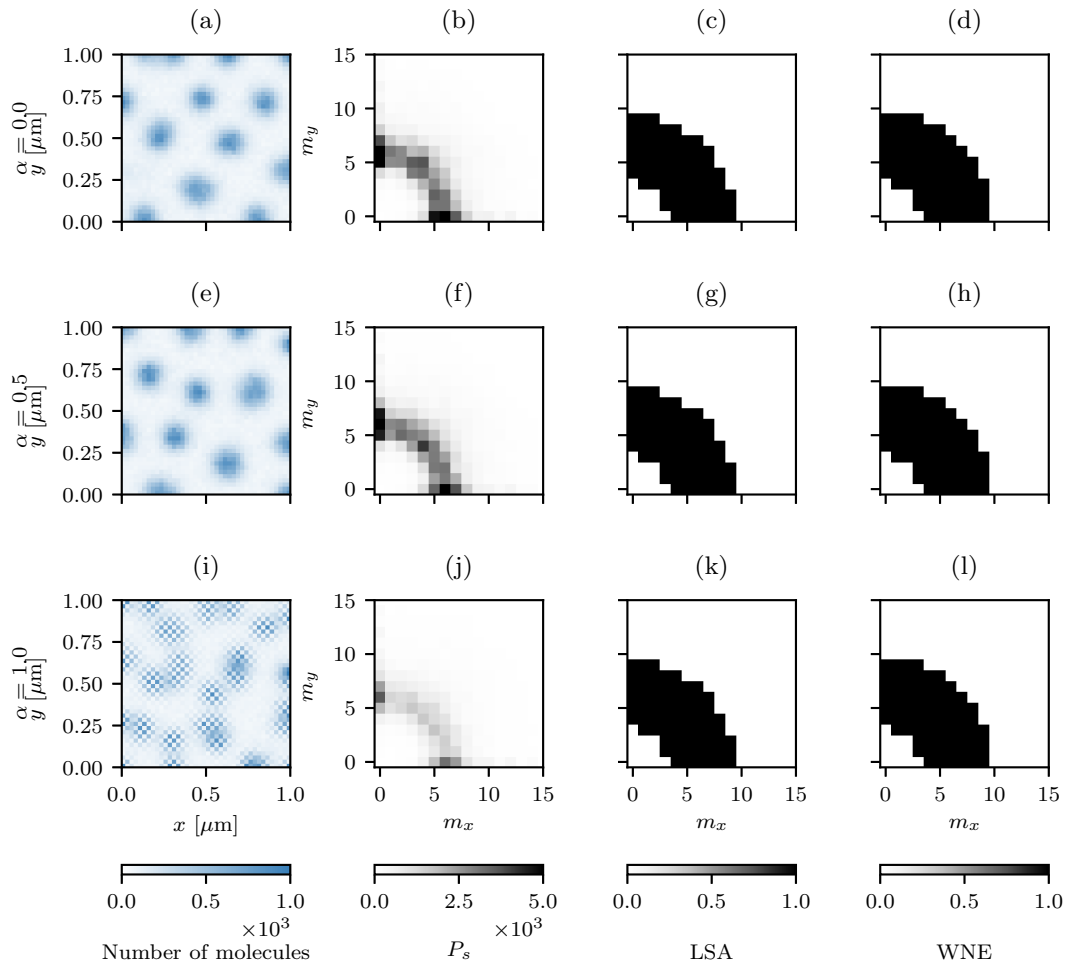}
\caption{Results from simulations with FDM-derived diffusive jump rates with $\kappa = 1.0$. (a), (e), (i) An example pattern. (b), (f), (j) Averaged power spectrum over 100 simulations. (c), (g), (k) Wavemodes predicted using LSA. (d), (h), (l) Wavemodes predicted using the WNE.}
\label{Fig:power_spectrum_fdm}
\end{figure}

The left-most column of Figure~\ref{Fig:power_spectrum_fdm} shows examples of patterns formed with FDM-derived diffusive jump rates, while the second-from-left column of Figure~\ref{Fig:power_spectrum_fdm} shows corresponding power spectra averaged over 100 simulations. There is good agreement with the possible wavemodes predicted using the WNE (right-most column of Figure~\ref{Fig:power_spectrum_fdm}) and LSA (second-from-right column of Figure~\ref{Fig:power_spectrum_fdm}). Our results indicate that varying the parameter $\alpha$, which controls the rate of diagonal jumping, has no discernible effect on pattern formation, except when $\alpha=1.0$ (only diagonal jumps are possible) where checkerboard patterns arise.


\pagebreak

\paragraph{The finite element method}

\begin{figure}[h]
\centering
\includegraphics[scale=1.0]{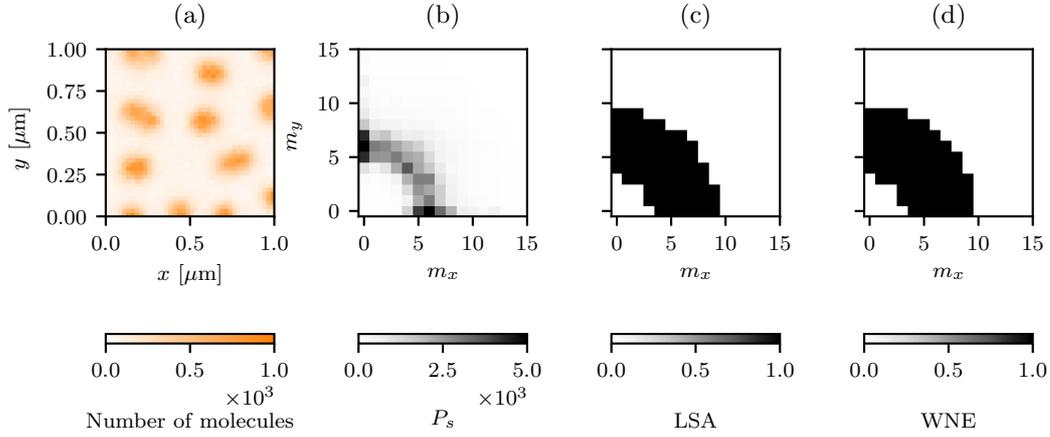}
\caption{Results from simulations with FEM-derived diffusive jump rates with $\kappa = 1.0$. (a) An example pattern. (b) Averaged power spectrum over 100 simulations. (c) Wavemodes predicted using LSA. (d) Wavemodes predicted using the WNE.}
\label{Fig:power_spectrum_fem}
\end{figure}

The left-most plot in Figure~\ref{Fig:power_spectrum_fem} shows an example pattern formed with FEM-derived diffusive jump rates, while the second-from-left plot shows the corresponding power spectra averaged over 100 simulations. There is good agreement with the possible wavemodes predicted using the WNE (right-most plot of Figure~\ref{Fig:power_spectrum_fem}) and LSA (second-from-right plot of Figure~\ref{Fig:power_spectrum_fem}).


\clearpage
\pagebreak

\paragraph{The finite volume method}

\begin{figure}[h]
\centering
\includegraphics[scale=1.0]{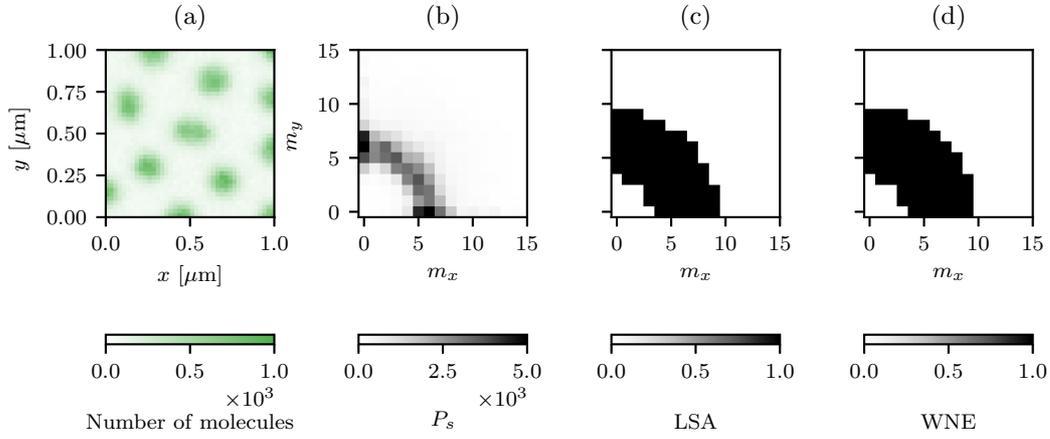}
\caption{Results from simulations with FVM-derived diffusive jump rates with $\kappa = 1.0$. (a) An example pattern. (b) Averaged power spectrum over 100 simulations. (c)  Wavemodes predicted using LSA. (d)  Wavemodes predicted using the WNE.}
\label{Fig:power_spectrum_fvm}
\end{figure}

For the case of FVM-derived diffusive jump rates, the simulations produced very similar results to the FEM-derived diffusive jump rates. The averaged power spectrum, as well as the predicted wavemodes using LSA and the WNE, are very similar for the FVM and FEM cases, as can be seen in Figure~\ref{Fig:power_spectrum_fvm}.


\pagebreak

\paragraph{The first exit time method}

\begin{figure}[h]
\centering
\includegraphics{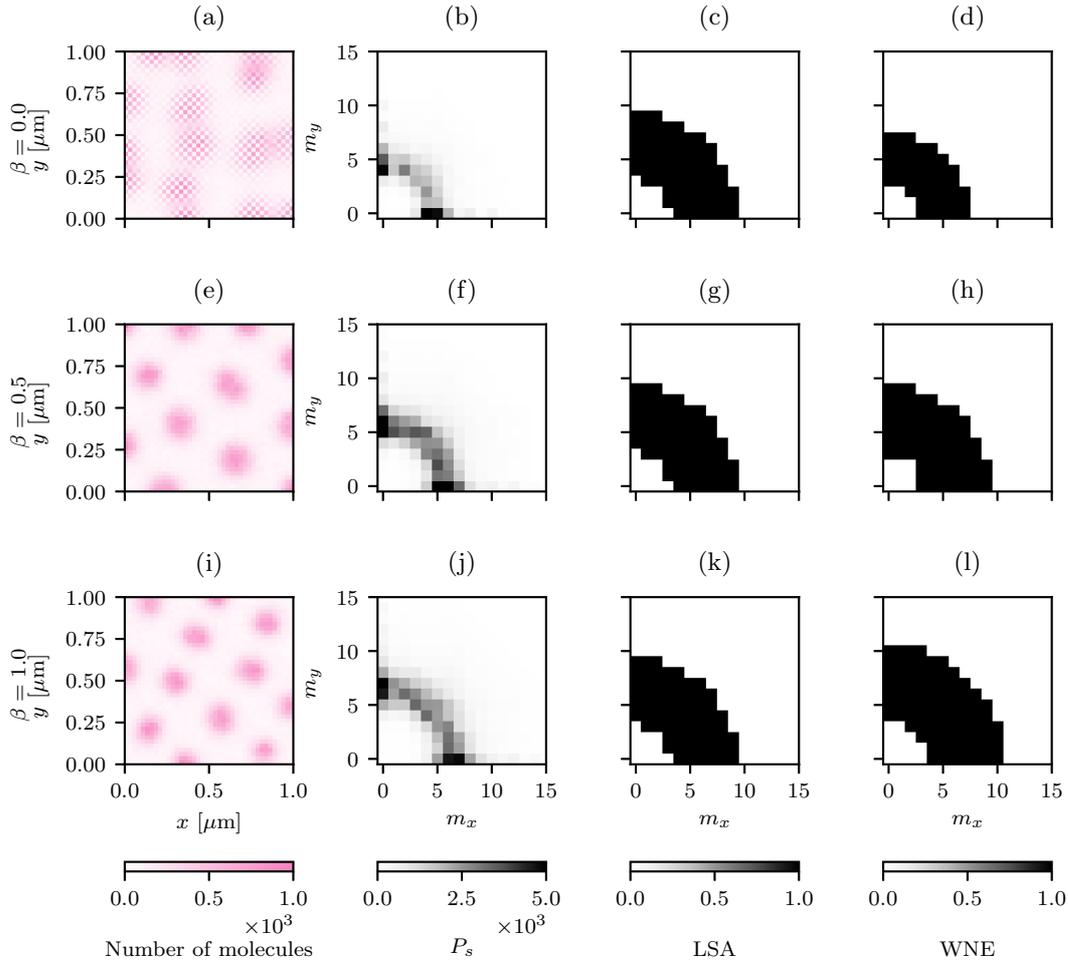}
\caption{Results from simulations with FET-derived diffusive jump rates with $\kappa = 1.0$. (a), (e), (i) An example pattern. (b), (f), (j) Averaged power spectrum over 100 simulations. (c), (g), (k) Wavemodes predicted using LSA. (d), (h), (l) Wavemodes predicted using the WNE. The chequerboard pattern seen in (a) is due to the jump rates being non-zero only in the diagonal directions when $\beta = 0$.}
\label{Fig:power_spectrum_fet_kappa_1_0}
\end{figure}

Lastly, we investigate the effects of using FET-derived diffusive jump rates on pattern formation. Based on the averaged power spectrum shown in Figure~\ref{Fig:power_spectrum_fet_kappa_1_0}(b), we see that changing the value of $\beta$ shifts the range of excited wavemodes towards larger values. This observation is consistent with the predictions made using the WNE (Figure~\ref{Fig:power_spectrum_fet_kappa_1_0}(c)). In this respect, the predictions of LSA fall down as they do not take into account details of the diffusive jump rates
(Figure~\ref{Fig:power_spectrum_fet_kappa_1_0}(d)).


\pagebreak

\begin{figure}[h]
\centering
\includegraphics{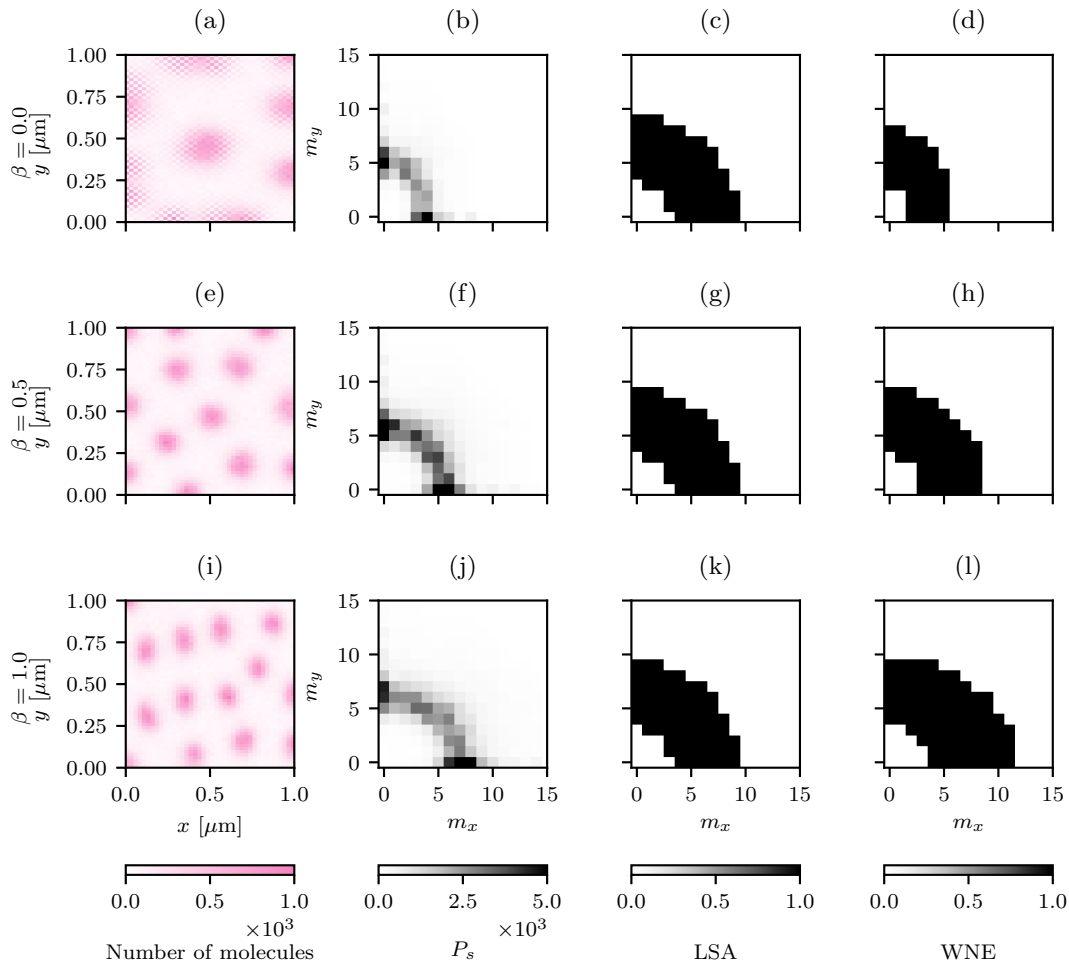}
\caption{Results from simulations with FET-derived diffusive jump rates with $\kappa = 1.4$. (a), (e), (i) An example pattern. (b), (f), (j) Averaged power spectrum over 100 simulations. (c), (g), (k) Wavemodes predicted using LSA. (d), (h), (l) Wavemodes predicted using the WNE. Again, we see checkerboard patterns for $\beta = 0$ where there are no diagonal jumps.}
\label{Fig:power_spectrum_fet_kappa_1_4}
\end{figure}

To explore whether changing the compartment aspect ratio has any bearing on the patterns formed, we also simulated the system with $\kappa = 1.4$. As in the case of $\kappa = 1.0$, larger wavemodes appear as the value of $\beta$ increases (Figure~\ref{Fig:power_spectrum_fet_kappa_1_4}(b)). However, the range of wavemodes observed in the $x$-direction is more significantly affected than the range in the $y$-direction. This observation is, again, consistent with the predictions made using the WNE, but cannot be predicted using LSA.


\enlargethispage{1cm}

In both cases, $\kappa = 1.0$ and $\kappa = 1.4$, changing the parameter $\beta$, which controls the rate of molecules jumping diagonally in the FET derivation, has a significant effect on the type of patterns formed, as shown in Figure~\ref{Fig:power_spectrum_fet_kappa_1_0} and Figure~\ref{Fig:power_spectrum_fet_kappa_1_4}.


\pagebreak

\subsection{Growing domain}
\label{Sec:growing_domain}

In this section, we explore the effects of using the different diffusive jump rate derivations on a growing domain.  As for the static domain case, we first consider cases where we can compare with the analytic solution of the corresponding macroscale PDE and we then consider models for pattern formation. We assume domain growth to be isotropic, such that $\Omega_t = [0, L(t)]^2$ and, for simplicity, we consider the domain to be growing exponentially in time so that
\begin{equation}
L(t) = L_0 e^{rt},
\end{equation}
where $L_0$ is the initial domain size and $r$ the growth rate.

As we now have time-dependent propensity functions, we cannot use the algorithm as stated in Section~\ref{Sec:model}. We adapt Algorithm \ref{Alg:SSA} using the Extrande method \cite{Voliotis2016}. As each of the propensity functions decreases with time, we use the propensity function calculated at time $t$ as the upper bound in the time interval, $[t, t+\delta t]$ in the Extrande method. Hence, the modified algorithm for SSA is as stated in Algorithm~\ref{Alg:SSA_for_growing}.


\begin{algorithm}
	\caption{Modified next sub-volume method SSA \label{Alg:SSA_for_growing}}
\vspace{1ex}
\begin{algorithmic}
	\State Set $t=0$.
	\State Initialise $\bm{u}$.
  \For{$i$ in $\left\{1,2,\ldots,I\right\}$}
  	\State Calculate the total propensity in compartment $i$, $a_i^0(\bm{u}(t), t)$.
	\State Set the upper bound $\tilde{a}_i(\bm{u}(t), t) = a_i^0(\bm{u}(t), t)$
    \State Initialise the time until the next reaction for compartment $i$: $t_i \sim \exp(\tilde{a}_i(\bm{u}(t), t))$.
  \EndFor
	\While{$t < T$}
    \State Find argmin of the set $\left\{t_1,t_2,\ldots,t_I\right\}$ and denote it $m$.
    \State Set $t = t_m$.
    \State Generate a random number $r \sim \textit{U}(0,1)$.
    \State Choose reaction / diffusion $q$ to fire. Note that reaction $k$ fires with probability $a_i^k(\bm{u}(t), t)/\tilde{a}_i(\bm{u}(t), t)$, diffusion of a molecule of species $\ell$ from box $i$ to box $j$ occurs with probability $d_{i,j}^{\,\ell}(\bm{u}(t), t)/\tilde{a}_i(\bm{u}(t), t)$ and no reaction happens with probability $( \tilde{a}_i(\bm{u}(t), t) - a_i^0(\bm{u}(t), t))/\tilde{a}_i(\bm{u}(t), t)$.
	\State Update molecule numbers: $\bm{u} := \bm{u} + \bm{\nu}_{q}$, where $\bm{\nu}_{q}$ is the stoichiometric matrix of reaction / diffusion $q$.
    \State Let $\Gamma$ be the set of indices of non-zero elements of $\bm{\nu}_q$.
    \For{$\gamma$ in $\Gamma$}
      \State Update $a_\gamma^k(\bm{u}(t), t)$, $d_{\gamma,j}^{\,\ell}(\bm{u}(t), t)$, $a_\gamma^0(\bm{u}(t), t)$ and $\tilde{a}_i(\bm{u}(t), t)$ accordingly.
      \State Generate a random number $\Delta t \sim \exp(\tilde{a}_\gamma(\bm{u}(t), t))$.
      \State Update the time until the next reaction: $t_{\gamma} = t + \Delta t$.
    \EndFor
	\EndWhile
\end{algorithmic}
\vspace{1ex}
\end{algorithm}

\clearpage

\subsubsection{Diffusion}
\label{Sec:results_first_order_growing}

As in the static case, there is an analytic solution to the macroscale PDE (see the Supplementary Information for details). Figure~\ref{Fig:error_against_time} shows how the error, as defined in Section~\ref{Sec:static_domain}, evolves in time. We can see that in all the cases shown in Figure~\ref{Fig:error_against_time}, whether for a static or a growing case, the error decreases with time and that the simulations with FVM-derived diffusive jump rates have the lowest error.


\begin{figure}[h]
\centering
\includegraphics{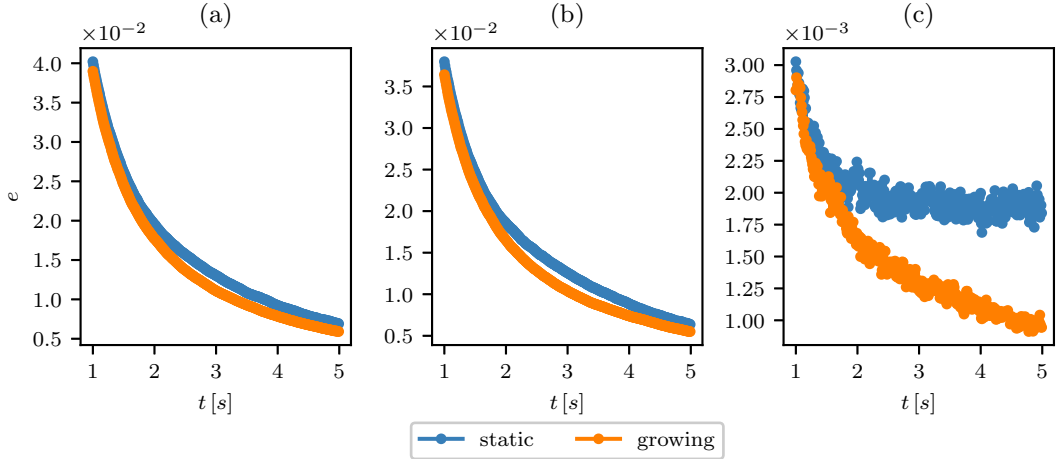}
\caption{Plots of the error, $e$ as a function of time for simulations with diffusion coefficient $D = 1  \, \mu m^2 s^{-1}$ and growth rate $r = 0.1 s^{-1}$. In each plot we compare the errors from simulations for a static domain and a growing domain where the jump rates were derived using: (a) the FDM ($\alpha = 0.9$), (b) the FEM, and (c) the FVM. The domain, of initial size $[0,5]\times[0,5]$, has been discretised into $21 \times 21$ compartments and all $5\times10^6$ molecules are initially placed into the bottom-right compartment.}
\label{Fig:error_against_time}
\end{figure}



\begin{figure}
\centering
\includegraphics{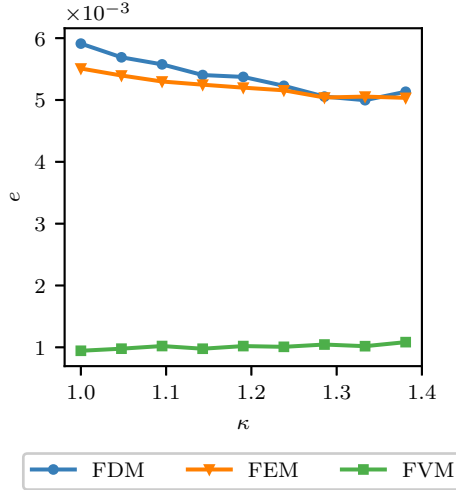}
\caption{Plots of the error, $e$, as a function of voxel aspect ratio, $\kappa$, for simulations with diffusion coefficient $D = 1  \, \mu m^2 s^{-1}$ and growth rate $r=0.1 \, s^{-1}$. We compare different derivations of the diffusive jump rates as we vary $\kappa$ For the FDM, $\alpha = 0.7$. The domain, of initial size $[0,5]\times[0,5]$, is discretised into $21 \times 21\kappa$ compartments and all $5\times10^6$ molecules are initially placed into the bottom-right compartment.}
\label{Fig:error_against_kappa_growing}
\end{figure}

Next, we explore the effects of changing the voxel aspect ratio, $\kappa$, on the error, presenting results in Figure~\ref{Fig:error_against_kappa_growing}. The FVM-derived diffusive jump rates provide the smallest error across the range of compartment aspect ratios considered, increasing slightly as the aspect ratio increases. On the other hand, for both the FEM- and FVM-derived diffusive jump rates, there is a slight decrease in the error as the compartment aspect ratio is increased.


\pagebreak


\subsubsection{Schnakenberg kinetics}
\label{Sec:results_schnakenberg_kinetics_growing}

\begin{figure}[h!]
\centering
\includegraphics{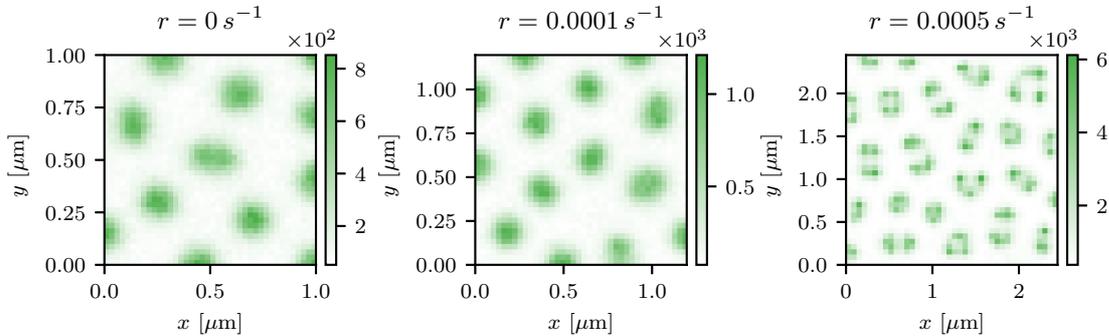}
\vspace{-1.5cm}
\caption{Comparison of different growth rates for simulations of the Schnakenberg system with the following parameters: $D_u = 1 \times 10^{-5} \, \mu m^2 s^{-1}$, $D_v = 0.001 \, \mu m^2 s^{-1}$, $k_1 = 1.0 \, \mu m^{-2} s^{-1}$, $k_2 = 0.02 \, s^{-1}$, $k_3 = 1\times10^{-6} \, \mu m^4 s^{-1}$, $k_4 = 3.0 \, \mu m^{-2} s^{-1}$ run until $T = 1800 \, s$ and FVM-derived jump rates used. The domain has been discretised into $40 \times 40$ compartments.}
\label{Fig:schnakenberg_growing_comparison}
\end{figure}

We now consider pattern formation on a growing domain to determine to what extent the different derivations of diffusive jump rates affect patterns formed. We again use the Schnakenberg system, as in Section~\ref{Sec:results_schnakenberg_kinetics}. Figure~\ref{Fig:schnakenberg_growing_comparison} showcases the range of patterns that form as the domain growth rate is varied.


\begin{figure}
\centering
\includegraphics{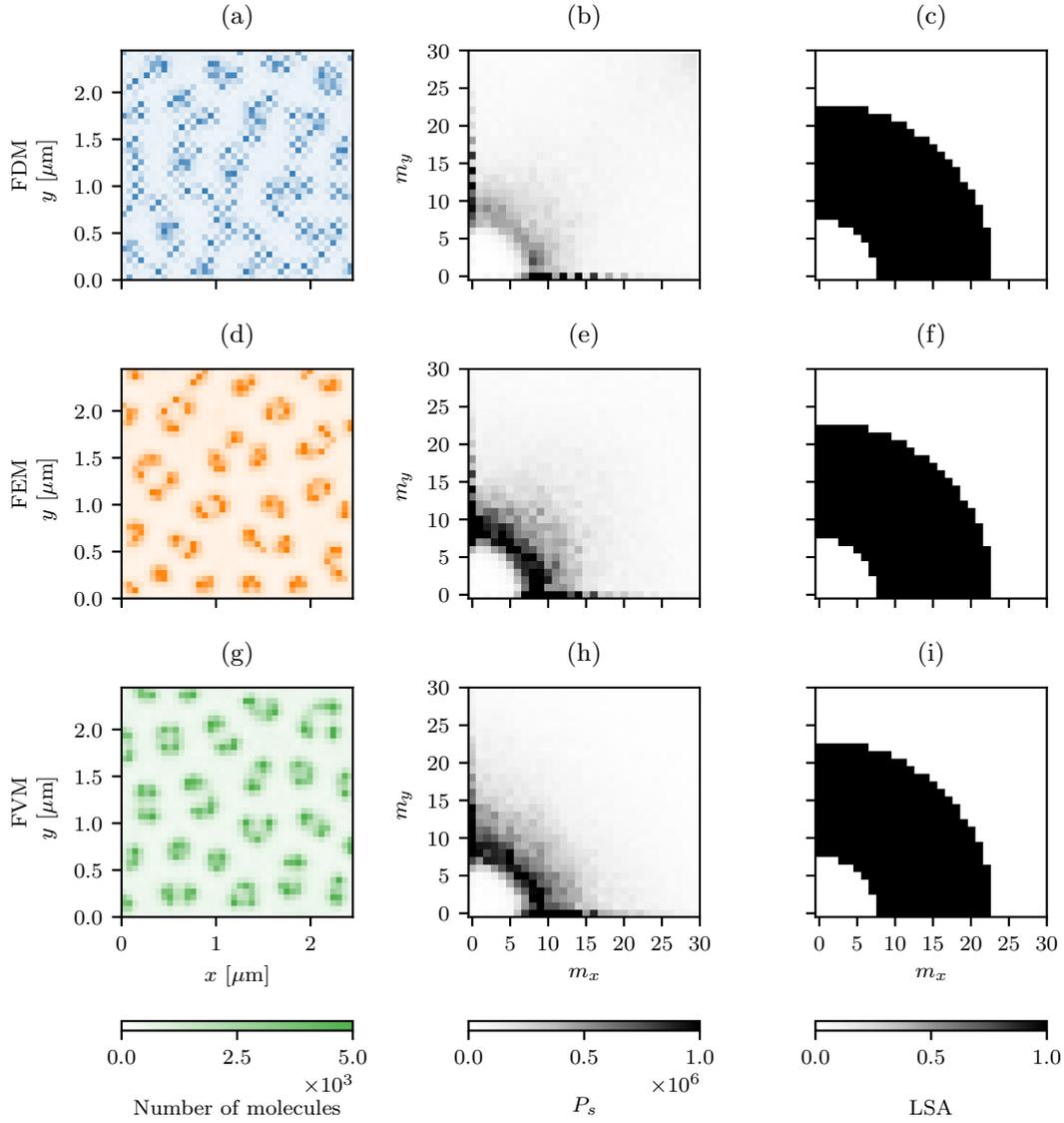}
\caption{Results from simulations of the Schnakenberg system on a growing domain with the following parameters: $D_u = 1.0 \times 10^{-5} \, \mu m^2 s^{-1}$, $D_v = 0.001 \, \mu m^2 s^{-1}$, $k_1 = 1.0 \, \mu m^{-2} s^{-1}$, $k_2 = 0.02 \, s^{-1}$, $k_3 = 1\times10^{-6} \, \mu m^{4} s^{-1}$, $k_4 = 3.0 \, \mu m^{-2} s^{-1}$ run until $T = 1800 \, s$. The domain has been discretised into $40 \times 40$ compartments. Left-hand column -- example patterns; centre column -- averaged power spectrum over 100 simulations; right-hand column --  predictions from LSA.}
\label{Fig:schnakenberg_growing_diff_rates}
\end{figure}

We see in Figure~\ref{Fig:schnakenberg_growing_diff_rates} that the patterns formed are different compared to the static case. However, using different derivations for the diffusive jump rates does not seem to markedly affect the patterns produced. The reason we see different patterns when the domain grows is because higher wavemodes become available, and the inherent stochasticity in the system causes clusters of molecules (peaks) to split and these higher wavemodes are then realised. This change in the wavemode is highlighted in the centre and left-hand columns in Figure~\ref{Fig:schnakenberg_growing_diff_rates}.


\clearpage

\section{Conclusions}
\label{Sec:conclusions}

Our aim in this work was to investigate the effects of different methods of derivation of diffusive jump rates on stochastic reaction-diffusion systems modelled using the RDME. To do so, we replicated, extended and applied the results of Meinecke and L\"otstedt~\cite{meinecke2016stochastic} to a number of reaction-diffusion systems, that included a range of reaction types, on both static and growing domains.

For the production-decay examples, which consisted solely of zeroth and first order reactions, the FVM generally gives the most reliable results in terms of the error between stochastic simulations and solution of the corresponding macroscale PDE (as does the FDM for $\alpha=0$ where there are no diagonal jumps and we have the same diffusive jump rates as for the FVM) (Figure~\ref{Fig:plot_of_error_against_kappa}). When bimolecular reactions are included, we showed that the spatial discretisation affects results, for example shifting the stationary distribution (Figure~\ref{Fig:stationary_dists}).

We then moved to study the effects of different methods of derivation of diffusive jump rates on pattern formation, using the Turing reaction-diffusion model with Schnakenberg kinetics. We compared the averaged power spectra from  simulations with the wavemodes predicted using LSA, which analyses the corresponding PDE model, and the WNE, which takes into account the details of the diffusive jump rates (Figures~\ref{Fig:power_spectrum_fdm}--\ref{Fig:power_spectrum_fet_kappa_1_0}). Our results show that the WNE approach provides good predictions of the range of possible wave modes, as expected from the work of Woolley \textit{et al.}~\cite{woolley2011power}. In particular, we showed that the aspect ratio of the compartments can significantly affect the patterns formed (see Figure~\ref{Fig:power_spectrum_fet_kappa_1_4}).

We also considered the effects of uniform domain growth on derivations of the diffusive jump rates, and simulations of pattern formation. We considered the FDM, FVM and FEM showing, in line with the results of Woolley \textit{et al.}~\cite{woolley2011power}, that the diffusive jump rates become time-dependent as the compartment dimensions now change in time. It is not possible to use the FET approach in the context of growing domains because the mean of the FET distribution becomes infinite, and it is difficult to approximate the FET distribution with an exponential distribution.

Taken together, the results presented in this work demonstrate that care should be taken when choosing how to model  diffusion in the context of RDME models of stochastic reaction-diffusion systems. Future work will consider how to extend these results to unstructured meshes and more complicated forms of domain growth.


\noindent
\textbf{Acknowledgements}

R.E.B. is a Royal Society Wolfson Research Merit Award holder, would like to thank the Leverhulme Trust for a Research Fellowship. B.J.B. would like to thank the EPSRC for supporting this research through grant EP/G03706X/1.


\bibliography{refs}
\bibliographystyle{abbrv}


\end{document}



\begin{frontmatter}

\title{Supplementary information: \\Effects of different discretisations of the Laplacian upon stochastic
simulations of reaction-diffusion systems \\ on both static and growing domains}%

\author{Bartosz J. Bartmanski$^\dagger$ and Ruth E. Baker$^\dagger$}

\address{$^\dagger$Mathematical Institute, University of Oxford, \\ Woodstock Road, Oxford, OX2 6GG, UK}


\begin{abstract}
By discretising space into compartments and letting system dynamics be governed by the reaction-diffusion master equation, it is possible to derive and simulate a stochastic model of reaction and diffusion on an arbitrary domain. However, there are many implementation choices involved in this process, such as the choice of discretisation and method of derivation of the diffusive jump rates, and it is not clear \textit{a priori} how these affect model predictions. To shed light on this issue, in this work we explore how a variety of discretisations and method for derivation of the diffusive jump rates affect the outputs of stochastic simulations of reaction-diffusion models, in particular using Turing's model of pattern formation as a key example. We consider both static and uniformly growing domains and demonstrate that, while only minor differences are observed for simple reaction-diffusion systems, there can be vast differences in model predictions for systems that include complicated reaction kinetics, such as Turing's model of pattern formation. Our work highlights that care must be taken in using the reaction-diffusion master equation to make predictions as to the dynamics of stochastic reaction-diffusion systems.
\end{abstract}


\begin{keyword}
stochastic simulation \sep diffusion \sep reaction \sep pattern formation.
\MSC[2010] 35K05 \sep 68U20 \sep 65C35 92-08.
\end{keyword}


\end{frontmatter}



\pagebreak


\section{Finite element method derivation}
\label{Sec_supp_FEM}

Here we derive the elements of the matrices $M$ and $N$ in the FEM derivation of jump rates. We have
\begin{equation}
	M \frac{\partial \tilde{\textbf{p}}}{\partial t} = - N \tilde{\textbf{p}},
\end{equation}
where $M_{ij} = \int_\Omega \varphi_i \varphi_j \D \omega$ and $N_{ij} = \int_\Omega \nabla \varphi_i \cdot \nabla \varphi_j \D \omega$. We wish to find the non-zero elements of the matrices $M$ and $N$. As we are
interested in deriving the jump rates for a non-boundary compartment, we will assume that the element (indexed by $i$) of matrices $M$ and $N$ is also not on the boundary of the FEM discretisation. Given that the function $\varphi_i$ is zero outside of its neighbourhood, the neighbourhood being $[x_i-\kappa h, x_i + \kappa h] \times [y_i - h, y_i + h]$, then we only have nine elements to calculate in both matrices $M$ and $N$ for the $i$th row. The nine elements in the $i$th row being due to the eight adjacent points and from itself.

We defined the function $\varphi_i$ as
\begin{equation}
	\varphi_i \left( x, y \right) =
	\left\{
	\begin{array}{cc}
		\left(1-\cfrac{x-x_i}{\kappa h}\right) \left(1-\cfrac{y-y_i}{h}\right)
		& \text{for} \quad \textbf{x} \in
		\left[ x_i , x_i + \kappa h \right] \times \left[ y_i , y_i + h \right] , \\
		\left(1+\cfrac{x-x_i}{\kappa h}\right) \left(1-\cfrac{y-y_i}{h}\right)
		& \text{for} \quad \textbf{x} \in
		\left[ x_i - \kappa h , x_i \right] \times \left[ y_i , y_i + h \right] , \\
		\left(1+\cfrac{x-x_i}{\kappa h}\right) \left(1+\cfrac{y-y_i}{h}\right)
		& \text{for} \quad \textbf{x} \in
		\left[ x_i - \kappa h , x_i \right] \times \left[ y_i - h , y_i \right]  , \\
		\left(1-\cfrac{x-x_i}{\kappa h}\right) \left(1+\cfrac{y-y_i}{h}\right)
		& \text{for} \quad \textbf{x} \in
		\left[ x_i, x_i + \kappa h \right] \times \left[ y_i - h , y_i \right]  , \\
		0 & \text{otherwise}  .
	\end{array}
	\right.
\end{equation}
Before we proceed with the calculations of the elements of matrices $M$ and $N$, we need to determine the gradient of the function $\varphi_i$, as the integrals in the elements of the matrix $N$ involve gradient of the function $\varphi$. The partial derivatives with respect to $x$ and $y$ are as follows
\begin{equation}
	\frac{\partial \varphi_i}{\partial x} \left( x, y \right) =
	\left\{
	\begin{array}{cc}
		-\frac{1}{\kappa h} \left(1-\cfrac{y-y_i}{h}\right)
		& \text{for} \quad \textbf{x} \in
		\left[ x_i , x_i + \kappa h \right] \times \left[ y_i , y_i + h \right]  , \\
		\frac{1}{\kappa h} \left(1-\cfrac{y-y_i}{h}\right)
		& \text{for} \quad \textbf{x} \in
		\left[ x_i - \kappa h , x_i \right] \times \left[ y_i , y_i + h \right]  , \\
		\frac{1}{\kappa h} \left(1+\cfrac{y-y_i}{h}\right)
		& \text{for} \quad \textbf{x} \in
		\left[ x_i - \kappa h , x_i \right] \times \left[ y_i - h , y_i \right]  , \\
		-\frac{1}{\kappa h} \left(1+\cfrac{y-y_i}{h}\right)
		& \text{for} \quad \textbf{x} \in
		\left[ x_i, x_i + \kappa h \right] \times \left[ y_i - h , y_i \right]  , \\
		0 & \text{otherwise},
	\end{array}
	\right.
\end{equation}
and
\begin{equation}
	\frac{\partial \varphi_i}{\partial y} \left( x, y \right) =
	\left\{
	\begin{array}{cc}
		-\frac{1}{h} \left(1-\cfrac{x-x_i}{\kappa h}\right)
		& \text{for} \quad \textbf{x} \in
		\left[ x_i , x_i + \kappa h \right] \times \left[ y_i , y_i + h \right]  , \\
		-\frac{1}{h} \left(1+\cfrac{x-x_i}{\kappa h}\right)
		& \text{for} \quad \textbf{x} \in
		\left[ x_i - \kappa h , x_i \right] \times \left[ y_i , y_i + h \right]  , \\
		\frac{1}{h} \left(1+\cfrac{x-x_i}{\kappa h}\right)
		& \text{for} \quad \textbf{x} \in
		\left[ x_i - \kappa h , x_i \right] \times \left[ y_i - h , y_i \right]  , \\
		\frac{1}{h} \left(1-\cfrac{x-x_i}{\kappa h}\right)
		& \text{for} \quad \textbf{x} \in
		\left[ x_i, x_i + \kappa h \right] \times \left[ y_i - h , y_i \right]  . \\
		0 & \text{otherwise} .
	\end{array}
	\right.
\end{equation}

We have discretised the domain $\Omega$, such that there are $n_x$ compartments in the $x$-direction and $n_y$ compartments in the $y$-direction. Hence, the index of the compartment above compartment $i$ is $i+n_x$ and
the index of the compartment below is $i-n_x$. For simplicity, let us denote $(1)$ to be the area
$\left[ x_i , x_i + \kappa h \right] \times \left[ y_i , y_i + h \right]$,
$(2)$ to be the area
$\left[ x_i - \kappa h , x_i \right] \times \left[ y_i , y_i + h \right]$,
$(3)$ to be the area
$\left[ x_i - \kappa h , x_i \right] \times \left[ y_i - h , y_i \right]$ and
$(4)$ to be the area
$\left[ x_i, x_i + \kappa h \right] \times \left[ y_i - h , y_i \right]$.

Now we can calculate the non-zero elements of row $i$ of matrix $M$, which are
\begin{eqnarray}
	\int_\Omega \varphi_i^2 \D \omega &=&
	\int_{(1)} \left(1-\frac{x-x_i}{\kappa h}\right) \left(1-\frac{y-y_i}{h}\right) \left(1-\frac{x-x_i}{\kappa h}\right) \left(1-\frac{y-y_i}{h}\right) \D x \D y + \nonumber \\
	&& \quad \int_{(2)} \left(1+\frac{x-x_i}{\kappa h}\right) \left(1-\frac{y-y_i}{h}\right) \left(1+\frac{x-x_i}{\kappa h}\right) \left(1-\frac{y-y_i}{h}\right) \D x \D y + \nonumber \\
	&& \quad \int_{(3)} \left(1+\frac{x-x_i}{\kappa h}\right) \left(1+\frac{y-y_i}{h}\right) \left(1+\frac{x-x_i}{\kappa h}\right) \left(1+\frac{y-y_i}{h}\right) \D x \D y + \nonumber \\
	&& \quad \int_{(4)} \left(1-\frac{x-x_i}{\kappa h}\right) \left(1+\frac{y-y_i}{h}\right) \left(1-\frac{x-x_i}{\kappa h}\right) \left(1+\frac{y-y_i}{h}\right) \D x \D y \nonumber  \\
	&=& \frac{4 \kappa h^2}{9}  ,
\end{eqnarray}
\begin{eqnarray}
	\int_\Omega \varphi_i \varphi_{i+1} \D \omega &=& \int_{(1)} \left(1-\frac{x-x_i}{\kappa h}\right) \left(1-\frac{y-y_i}{h}\right) \left(\frac{x-x_i}{\kappa h}\right) \left(1-\frac{y-y_i}{h}\right) \D x \D y + \nonumber \\
	&& \quad \int_{(4)} \left(1-\frac{x-x_i}{\kappa h}\right) \left(1+\frac{y-y_i}{h}\right) \left(\frac{x-x_i}{\kappa h}\right) \left(1+\frac{y-y_i}{h}\right) \D x \D y \nonumber  \\
	&=& \frac{\kappa h^2}{9} ,
\end{eqnarray}
\begin{eqnarray}
	\int_\Omega \varphi_i \varphi_{i+n_x+1} \D \omega &=& \int_{(1)} \left(1-\frac{x-x_i}{\kappa h}\right) \left(1-\frac{y-y_i}{h}\right) \left(\frac{x-x_i}{\kappa h}\right) \left(\frac{y-y_i}{h}\right) \D x \D y \nonumber \\
	&=& \frac{\kappa h^2}{36},
\end{eqnarray}
\begin{eqnarray}
		\int_\Omega \varphi_i \varphi_{i+n_x} \D \omega &=& \int_{(1)} \left(1-\frac{x-x_i}{\kappa h}\right) \left(1-\frac{y-y_i}{h}\right) \left(1-\frac{x-x_i}{\kappa h}\right) \left(\frac{y-y_i}{h}\right) \D x \D y + \nonumber \\
	&& \quad \int_{(2)} \left(1+\frac{x-x_i}{\kappa h}\right) \left(1-\frac{y-y_i}{h}\right) \left(1+\frac{x-x_i}{\kappa h}\right) \left(\frac{y-y_i}{h}\right) \D x \D y \nonumber  \\
	&=& \frac{\kappa h^2}{9} .
\end{eqnarray}
By symmetry of the discretisation the other elements in the $i$th row of matrix $M$ are
\begin{eqnarray}
	\int_\Omega \varphi_i \varphi_{i-1} \D \omega &=& \frac{\kappa h^2}{9} , \\
	\int_\Omega \varphi_i \varphi_{i-n_x} \D \omega &=& \frac{\kappa h^2}{9}  , \\
	\int_\Omega \varphi_i \varphi_{i+n_x-1} \D \omega &=& \frac{\kappa h^2}{36}  , \\
	\int_\Omega \varphi_i \varphi_{i-n_x-1} \D \omega &=& \frac{\kappa h^2}{36}  , \\
	\int_\Omega \varphi_i \varphi_{i-n_x+1} \D \omega &=& \frac{\kappa h^2}{36}  .
\end{eqnarray}
As we will use lumped mass matrix method \cite{Wu2006} on the matrix $M$,
the resulting matrix $\tilde{M}$ will be a diagonal matrix,
with the non-zero element in its $i$th row being the sum of
all elements in that row. Therefore, $i$th row non-zero
element of matrix $\tilde{M}$ is $\kappa h^2$.

We still need to calculate all the non-zero elements of matrix $N$. Using the same arguments as when calculating the non-zero elements of matrix $M$, we need only calculate four integrals, specifically
\begin{eqnarray}
	\int_\Omega \left| \nabla \varphi_i \right|^2 \D \omega &=& \int_{(1)} \left(-\frac{1}{\kappa h}\right) \left(1-\frac{y-y_i}{h}\right) \left(-\frac{1}{\kappa h}\right) \left(1-\frac{y-y_i}{h}\right) \D x \D y + \nonumber \\
	&& \quad \int_{(1)} \left(1-\frac{x-x_i}{\kappa h}\right) \left(-\frac{1}{h}\right) \left(1-\frac{x-x_i}{\kappa h}\right) \left(-\frac{1}{h}\right) \D x \D y + \nonumber \\
	&& \quad \int_{(2)} \left(\frac{1}{\kappa h}\right) \left(1-\frac{y-y_i}{h}\right) \left(\frac{1}{\kappa h}\right) \left(1-\frac{y-y_i}{h}\right) \D x \D y + \nonumber \\
	&& \quad \int_{(2)} \left(1+\frac{x-x_i}{\kappa h}\right) \left(-\frac{1}{h}\right) \left(1+\frac{x-x_i}{\kappa h}\right) \left(-\frac{1}{h}\right) \D x \D y + \nonumber \\
	&& \quad \int_{(3)} \left(\frac{1}{\kappa h}\right) \left(1+\frac{y-y_i}{h}\right) \left(\frac{1}{\kappa h}\right) \left(1+\frac{y-y_i}{h}\right) \D x \D y + \nonumber \\
	&& \quad \int_{(3)} \left(1+\frac{x-x_i}{\kappa h}\right) \left(\frac{1}{h}\right) \left(1+\frac{x-x_i}{\kappa h}\right) \left(\frac{1}{h}\right) \D x \D y + \nonumber \\
	&& \quad \int_{(4)} \left(-\frac{1}{\kappa h}\right) \left(1+\frac{y-y_i}{h}\right) \left(-\frac{1}{\kappa h}\right) \left(1+\frac{y-y_i}{h}\right) \D x \D y + \nonumber \\
	&& \quad \int_{(4)} \left(1-\frac{x-x_i}{\kappa h}\right) \left(\frac{1}{h}\right) \left(1-\frac{x-x_i}{\kappa h}\right) \left(\frac{1}{h} \right) \D x \D y \nonumber \\
	&=& \frac{4 \kappa}{3} + \frac{4}{3 \kappa} ,
\end{eqnarray}
\begin{eqnarray}
	\int_\Omega (\nabla \varphi_i) \cdot (\nabla \varphi_{i+1}) \D \omega &=& \int_{(1)} \left(-\frac{1}{\kappa h}\right) \left(1-\frac{y-y_i}{h}\right) \left(\frac{1}{\kappa h}\right) \left(1-\frac{y-y_i}{h}\right) \D x \D y + \nonumber \\
	&& \quad \int_{(1)} \left(1-\frac{x-x_i}{\kappa h}\right) \left(-\frac{1}{h}\right) \left(\frac{x-x_i}{\kappa h}\right) \left(-\frac{1}{h}\right) \D x \D y + \nonumber \\
	&& \quad \int_{(4)} \left(-\frac{1}{\kappa h}\right) \left(1+\frac{y-y_i}{h}\right) \left(\frac{1}{\kappa h}\right) \left(1+\frac{y-y_i}{h}\right) \D x \D y \nonumber  \\
	&& \quad \int_{(4)} \left(1-\frac{x-x_i}{\kappa h}\right) \left(\frac{1}{h}\right) \left(\frac{x-x_i}{\kappa h}\right) \left(\frac{1}{h}\right) \D x \D y \nonumber  \\
	&=& \frac{\kappa}{3} - \frac{2}{3\kappa},
\end{eqnarray}
\begin{eqnarray}
	\int_\Omega (\nabla \varphi_i) \cdot (\nabla \varphi_{i+n_x+1}) \D \omega &=& \int_{(1)} \left(-\frac{1}{\kappa h}\right) \left(1-\frac{y-y_i}{h}\right) \left(\frac{1}{\kappa h}\right) \left(\frac{y-y_i}{h}\right) \D x \D y + \nonumber \\
	&& \quad \int_{(1)} \left(-\frac{1}{\kappa h}\right) \left(1-\frac{y-y_i}{h}\right) \left(\frac{1}{\kappa h}\right) \left(\frac{y-y_i}{h}\right) \D x \D y \nonumber \\
	&=& -\frac{\kappa}{6} -\frac{1}{6\kappa} ,
\end{eqnarray}
\begin{eqnarray}
		\int_\Omega ( \nabla \varphi_i) \cdot (\nabla \varphi_{i+n_x}) \D \omega &=& \int_{(1)} \left(-\frac{1}{\kappa h}\right) \left(1-\frac{y-y_i}{h}\right) \left(-\frac{1}{\kappa h}\right) \left(\frac{y-y_i}{h}\right) \D x \D y + \nonumber \\
	&& \quad \int_{(1)} \left(1-\frac{x-x_i}{\kappa h}\right) \left(-\frac{1}{h}\right) \left(1-\frac{x-x_i}{\kappa h}\right) \left(\frac{1}{h}\right) \D x \D y + \nonumber \\
	&& \quad \int_{(2)} \left(\frac{1}{\kappa h}\right) \left(1-\frac{y-y_i}{h}\right) \left(\frac{1}{\kappa h}\right) \left(\frac{y-y_i}{h}\right) \D x \D y + \nonumber  \\
	&& \quad \int_{(2)} \left(1+\frac{x-x_i}{\kappa h}\right) \left(-\frac{1}{h}\right) \left(1+\frac{x-x_i}{\kappa h}\right) \left(\frac{1}{h}\right) \D x \D y \nonumber  \\
	&=& \frac{1}{3\kappa} - \frac{2\kappa}{3} .
\end{eqnarray}
Hence, by symmetry, the other non-zero terms of row $i$ are
\begin{eqnarray}
	\int_\Omega (\nabla \varphi_i) \cdot ( \nabla \varphi_{i-1}) \D \omega &=& \frac{\kappa}{3} - \frac{2}{3\kappa}  , \\
	\int_\Omega (\nabla \varphi_i) \cdot ( \nabla \varphi_{i-n_x}) \D \omega &=& \frac{1}{3\kappa} - \frac{2\kappa}{3}  , \\
	\int_\Omega (\nabla \varphi_i) \cdot ( \nabla \varphi_{i+n_x-1}) \D \omega &=& -\frac{\kappa}{6} -\frac{1}{6\kappa}  , \\
	\int_\Omega (\nabla \varphi_i) \cdot ( \nabla \varphi_{i-n_x-1}) \D \omega &=& -\frac{\kappa}{6} -\frac{1}{6\kappa}  , \\
	\int_\Omega (\nabla \varphi_i) \cdot ( \nabla \varphi_{i-n_x+1}) \D \omega &=& -\frac{\kappa}{6} -\frac{1}{6\kappa} .
\end{eqnarray}
Using the above non-zero elements of the matricies $M$ and $N$, we can calculate
the jump rates for the FEM case.


\section{First exit time method derivation}
\label{Sec_supp_fet}

To derive the jump rates using the FET method, we first have to solve
\begin{equation}
	\frac{\p p}{\p t} =
	D \Delta p \quad \text{for} \quad \textbf{x} \in \omega_{i,j}  .
\end{equation}
Due to the translational invariance of this partial differential equation (PDE), we can solve it on the domain
$[0, 2\kappa h]\times[0, 2h]$, which has the following solution,
when solved using separation of variables,
\begin{equation}
\begin{split}
	p(x, y, t) =
	& \frac{1}{\kappa h^2} \sum_{k=1}^\infty \sum_{j=1}^\infty (-1)^{j+k}
	\sin \left(\frac{(2k-1)\pi x}{2\kappa h} \right)
	\sin \left(\frac{(2j-1)\pi y}{2 h} \right) \\
	& \qquad \qquad \qquad \qquad \times
	\exp\left(-\frac{\pi^2 D t}{4\kappa^2 h^2} \left( \kappa^2 (2j-1)^2 + (2k-1)^2 \right)\right)  .
\end{split}
\label{Eq:analytic_sol}
\end{equation}
From this solution we can calculate the survival probability using the equation
\begin{equation}
	S(t) = \mathbb{P}\left[ \tau \geqslant t \right] =
	\int_{\omega_{i,j}} p(\textbf{x}, t) \, \D \omega  .
\end{equation}
Hence, we have
\begin{equation}
\begin{split}
	S (t) &=
	\int_0^{2\kappa h } \int_0^{2h} p(\textbf{x}, t) \, \D x \D y \\
	&= \frac{1}{\kappa h^2} \int_0^{2\kappa h } \int_0^{2h}
	\sum_{k=1}^\infty \sum_{j=1}^\infty
	\sin \left( \frac{(2k-1) \pi x}{2\kappa h} \right)
	\sin \left( \frac{(2j-1) \pi y}{2 h} \right) \\
	& \qquad \cdot
	\exp\left(-\frac{\pi^2 D t}{4\kappa^2 h^2} \left(\kappa^2 (2j-1)^2 + (2k-1)^2 \right)\right) \\
	&= \frac{16}{\pi^2} \sum_{k=1}^\infty \sum_{j=1}^\infty
	\frac{(-1)^{j+k}}{(2k-1)(2j-1)}
	\exp\left(-\frac{\pi^2 D t}{4\kappa^2 h^2} \left(\kappa^2 (2j-1)^2 + (2k-1)^2 \right)\right)  .
\end{split}
\end{equation}
To calculate $\lambda_0$ we use Equations~(56) and~(57) from the main text to arrive at
\begin{equation}
	\lambda_0 =
	\dfrac{\pi^4 D}{64 \kappa^2 h ^2
	\mathlarger{\sum_{k=1}^\infty} \mathlarger{\sum_{j=1}^\infty}
	\dfrac{(-1)^{j+k}}{(2k-1)(2j-1)(\kappa^2 (2j-1)^2 + (2k-1)^2)}}  .
\end{equation}
To calculate the probabilities of jumping in each direction, $\theta_n$, we use Equation~\eqref{Eq:analytic_sol} and Equation~(60) from the main text to derive
\begin{eqnarray}
	\theta_1 &=&
	\sum_{k=1}^\infty \sum_{j=1}^\infty
	\frac{8(-1)^{k+1}}{\pi^2 (\kappa^2 (2j-1)^2 + (2k-1)^2))}
	\frac{2k-1}{2j-1} \sin \left(\frac{(2j-1)\pi\beta}{2} \right)  ,
	\label{Eq:FET_theta_1} \\
	\theta_2 &=&
	\sum_{k=1}^\infty \sum_{j=1}^\infty
	\frac{4(-1)^{j+k}}{\pi^2 (\kappa^2 (2j-1)^2 + (2k-1)^2))}
	\left\lbrace \frac{2k-1}{2j-1} \left(
	\sin \left( \pi (\beta j - \frac{\beta}{2} + j) \right) + 1
	\right) + \right. \nonumber \\
	&& \qquad \qquad \left. \frac{2j-1}{\kappa^2 (2k-1)} \left(
	\sin \left( \pi (\beta k - \frac{\beta}{2} + k) \right) + 1
	\right) \right\rbrace \, ,
	\label{Eq:FET_theta_2} \\
	\theta_3 &=&
	\sum_{k=1}^\infty \sum_{j=1}^\infty
	\frac{8(-1)^{j+1} \kappa^2}{\pi^2 (\kappa^2 (2j-1)^2 + (2k-1)^2))}
	\frac{2j-1}{2k-1} \sin \left(\frac{(2k-1)\pi\beta}{2}
	\right)  .
	\label{Eq:FET_theta_3}
\end{eqnarray}


\section{Analytic solutions}
\label{Sec_supp_analytic_solutions}

To compare the effects of using different methods of derivation of the diffusive jump rates, we compare the results of stochastic simulations with a large number of molecules with analytic solutions of the PDEs stated in the main text. In this section we shall state the analytical solutions we used for those comparisons. The derivations of these analytic solutions simply requires the use of separation of variables, hence, we omit the derivations. Furthermore, we denote
\begin{equation}
u_0 = 5 \times 10^6 / A,
\end{equation}
such that the initial condition used in many instances in the main text can be written as follows
\begin{equation}
    u(\textbf{x}, 0) =
    \left\{
    \begin{array}{cc}
        u_0 & \text{for } 0 < x < \kappa h, 0 < y < h, \\
        0 & \text{otherwise}.
    \end{array}
    \right.
\end{equation}


\subsection{Static domain}
\label{Subsec_supp_static_domain}

First, we state the analytic solution to Equation~(71) in the main text:
\begin{eqnarray}
\nonumber
u(\textbf{x}, t) &=& \frac{u_0 \kappa h^2}{L^2} \\
&& + \sum_{n=1}^\infty \frac{2 u_0 h}{n \pi L} \sin \left( \frac{n\pi \kappa h}{L} \right) \cos \left( \frac{n \pi x}{L} \right) \exp \left(-D t \left(\frac{n \pi}{L}\right)^2 \right) \nonumber\\
&& + \sum_{m=1}^\infty \frac{2 u_0 \kappa h}{m \pi L} \sin \left( \frac{m\pi h}{L} \right) \cos \left( \frac{m \pi y}{L} \right) \exp \left(-D t \left(\frac{m \pi}{L}\right)^2 \right) \nonumber\\
&& + \sum_{n=1}^\infty \sum_{m=1}^\infty \frac{4 u_0}{n m \pi^2 L^2} \sin \left( \frac{n\pi \kappa h}{L} \right) \sin \left( \frac{m\pi h}{L} \right) \cos \left( \frac{m \pi x}{L} \right) \cos \left( \frac{n \pi x}{L} \right) \nonumber\\
&& \qquad\qquad\qquad \times \exp \left(-D t \left(\frac{(m^2 + n^2)\pi^2}{L^2}\right) \right).
\label{Eq:diff_eq_analytic}
\end{eqnarray}
Next, we state the analytic solutions to Equations~(78)--(80) in the main text. We shall use $\tilde{u}(\textbf{x}, t)$ to denote the analytical solution stated above, \textit{i.e.} Equation~\eqref{Eq:diff_eq_analytic}. Then the analytical solution to Equation~(78) in the main text is
\begin{equation}
    u(\textbf{x}, t) = \tilde{u}(\textbf{x}, t) + k_1 t ,
\end{equation}
and the solution to Equation~(79) in the main text is
\begin{equation}
    u(\textbf{x}, t) = \tilde{u}(\textbf{x}, t) e^{k_{-1} t} ,
\end{equation}
and, finally, the analytical solution to Equation~(80) in the main text is
\begin{equation}
    u(\textbf{x}, t) = \tilde{u}(\textbf{x}, t) e^{k_{-1} t} + \frac{k_1}{k_{-1}} \left( 1 - e^{k_{-1} t} \right).
\end{equation}
When evaluating any of the formulas above numerically, we truncate the sum to the first million terms.


\subsection{Growing domain}
\label{Subsec_supp_growing_domain}

As we have calculated the analytical solution for the static domain case, so we must do the same for the growing domain case. In Section~3.2.1 of the main text we compare the results of the stochastic simulations with the analytical solution. As before, for brevity, we simply state the analytical solution (for derivation, we direct the reader to \cite{simpson2015exactcalc})
\begin{align}
\begin{split}
&u(\textbf{x}, t) = \frac{u_0 \kappa h^2}{L_0^2} \exp \left( -2rt \right) \\
& + \sum_{n=1}^\infty \frac{2 u_0 h}{n \pi L_0} \sin \left( \frac{n\pi \kappa h}{L_0} \right) \cos \left( \frac{n \pi x}{L_0} \right) \exp \left(-\frac{D}{2r} \left(\frac{n \pi}{L_0}\right)^2 \left( 1 - \exp (-2rt) \right) - 2rt \right) \\
& + \sum_{m=1}^\infty \frac{2 u_0 \kappa h}{m \pi L_0} \sin \left( \frac{m\pi h}{L_0} \right) \cos \left( \frac{m \pi y}{L_0} \right) \left(-\frac{D}{2r} \left(\frac{m \pi}{L_0}\right)^2 \left( 1 - \exp (-2rt) \right) - 2rt \right) \\
& + \sum_{n=1}^\infty \sum_{m=1}^\infty \frac{4 u_0}{n m \pi^2 L_0^2} \sin \left( \frac{n\pi \kappa h}{L_0} \right) \sin \left( \frac{m\pi h}{L_0} \right) \cos \left( \frac{m \pi x}{L_0} \right) \cos \left( \frac{n \pi x}{L_0} \right)\\
& \quad \quad \quad \times \exp \left(-\frac{D}{2r} \frac{(n^2+m^2) \pi^2}{L_0^2} \left( 1 - \exp (-2rt) \right) - 2rt \right) .
\end{split}
\label{Eq:diff_eq_analytic_growing}
\end{align}


\section{Linear stability analysis}
\label{Sec_supp_linear_stability_analysis}

An important tool in determining the spatial distribution of species of molecules in a deterministic reaction-diffusion pattern formation model is linear stability analysis (LSA). We outline the derivation of the conditions for spatial patterns to form for general reaction-diffusion kinetics, closely following the derivation outlined by Murray~\cite{murray2001mathematical}. Consider a PDE description of a two-species reaction-diffusion system
\begin{eqnarray}
	\frac{\p u}{\p t} \left( \textbf{x}, t \right) &=&
	D_u \frac{\p^2 u}{\p x^2} \left( \textbf{x}, t \right) + f(u,v) ,
	\label{Eq:general_system_equations_I} \\
	\frac{\p v}{\p t} \left( \textbf{x}, t \right) &=&
	D_v \frac{\p^2 v}{\p x^2} \left( \textbf{x}, t \right) + g(u,v)  ,
	\label{Eq:general_system_equations_II}
\end{eqnarray}
where $u$ is the concentration of species U and $v$ is the concentration of species V on the domain $\Omega$.
The initial conditions for this system are
\begin{equation}
	u(\textbf{x}, 0) = u_0 (\textbf{x})
	\quad \text{and} \quad
	v(\textbf{x}, 0) = v_0 (\textbf{x}) \, .
\end{equation}
We continue to use Neumann boundary conditions for this system, as in the rest of this paper. Let $\bar{u}$ and $\bar{v}$ be the homogeneous steady states of this system, \textit{i.e.} the values of $u$ and $v$ that satisfy the following equalities $f(\bar{u}, \bar{v}) = 0$ and $g(\bar{u}, \bar{v}) = 0$.

We linearise about the spatially uniform steady state by letting $u = \bar{u} + \delta u$ and $v = \bar{v} + \delta v$, where $\left| \delta u \right|, \left| \delta v \right| \ll 1$.
This leads to the following system of equations
\begin{equation}
	\frac{\p \textbf{w}}{\p t} = D \frac{\p^2 \textbf{w}}{\p x^2} + A \textbf{w}  ,
	\label{Eq:linear_stability_analysis_main}
\end{equation}
where $\textbf{w} = \left( \delta u, \delta v \right)^T$. Let us denote
\begin{equation}
	\left. \frac{\p f}{\p u} \right|_{(\bar{u}, \bar{v})} = f_u  ,
	\quad \left. \frac{\p f}{\p v} \right|_{(\bar{u}, \bar{v})} = f_v  ,
	\quad 	\left. \frac{\p g}{\p u} \right|_{(\bar{u}, \bar{v})} = g_u  ,
	\quad \left. \frac{\p g}{\p v} \right|_{(\bar{u}, \bar{v})} = g_v  ,
\end{equation}
so that we can write the matrices in Equation~\eqref{Eq:linear_stability_analysis_main} as
\begin{equation}
A =
\begin{bmatrix}
	f_u & f_v \\
	g_u & g_v
\end{bmatrix}
\quad \text{and} \quad
D =
\begin{bmatrix}
	D_u & 0 \\
	0 & D_v
\end{bmatrix} .
\end{equation}

Now, let us consider the system without diffusion, i.e.
\begin{equation}
	\frac{\p \textbf{w}}{\p t} = A \textbf{w} ,
	\label{Eq:no_diffusion_ODE_system}
\end{equation}
which is a linear system of ordinary differential equations. As a result, the solution is a linear superposition of solutions of the following type
\begin{equation}
	\textbf{w} = \textbf{c} e^{\lambda t} .
\end{equation}
The vector $\textbf{c}$ is a constant vector and $\lambda$ is a constant to be determined. If we now substitute the above solution into Equation~\eqref{Eq:no_diffusion_ODE_system}, then
\begin{equation}
	\lambda \textbf{w} = A \textbf{w},
\end{equation}
which has a non-trivial solution if and only if $\left| A - \lambda I \right| = 0$. Therefore, we can find the value of $\lambda$ to be
\begin{equation}
	\lambda_\pm = \frac{1}{2} \left[ f_u + g_v \pm \sqrt{(f_u + g_v)^2 - 4 (f_u g_v - f_v g_u)} \right] .
\end{equation}
As stated in \cite{murray2001mathematical}, we seek patterns driven by diffusion and, as a result, we require that the homogeneous solution be linearly stable, and hence $\text{Re}(\lambda_\pm) < 0$. Therefore, the first two conditions necessary for diffusion-driven instability are
\begin{equation}
	f_u + g_v < 0 \quad \text{and} \quad
	f_u g_v - f_v g_u > 0 .
	\label{Eq:conditions_for_patterns}
\end{equation}

We now consider the system with diffusion present and assume that the solution is of the form
\begin{equation}
	\textbf{w} = \sum_k \textbf{c}_k e^{\lambda t} \textbf{W}_k (\textbf{x}) \, ,
	\label{Eq:linear_stability_expression}
\end{equation}
where $\textbf{W}_k$ satisfies the eigenvalue problem
\begin{equation}
	\Delta W_k + k^2 W_k = 0 \quad \text{with} \quad ( \textbf{n} \cdot \nabla )
	\textbf{W}_k = 0 \quad \text{for} \quad \textbf{x} \in \p \Omega .
\end{equation}
Here, $\textbf{n}$ is the vector normal to the domain boundary, $\p \Omega$. If we now substitute Equation~\eqref{Eq:linear_stability_expression} into Equation~\eqref{Eq:linear_stability_analysis_main}, then we have
\begin{equation}
	0 = ( A - k^2 D - \lambda I) \textbf{W}_k  ,
\end{equation}
which has non-trivial solutions if and only if
\begin{equation}
	\left| A - k^2 D -\lambda I \right| = 0  .
\end{equation}
The above can be simplified to
\begin{equation}
	\lambda^2 + ( (D_u + D_v) k^2 - (f_u + g_v) ) \lambda + h(k^2) = 0  ,
\end{equation}
where $h(k^2) = D_u D_v k^4 - (D_u g_v + D_v f_u ) k^2 + f_u g_v - f_v g_u$. Since we require spatial patterns to form, we need linear instability, \textit{i.e.} $\text{Re} (\lambda) > 0$ for $k > 0$. This means that the system is
linearly unstable for the values of $k$ at which $h(k^2) > 0$. Therefore, given details of $\Omega$ we can find which wavenumbers, $(k_x, k_y)$, where $k^2 = k_x^2 +k_y^2$, that can be excited in any given system~\cite{murray2001mathematical}.


\section{Weak noise expansion}
\label{Sec_supp_weak_noise_expansion}

An alternative approach to LSA is to use a method based on the weak-noise expansion (WNE)~\cite{vanKampen2007}, as developed by Woolley \textit{et al.}~\cite{woolley2011power}. We use the two-dimensional spatial, discrete Fourier cosine transform
\begin{equation}
	\hat{f} (k_x, k_y) =
	\Delta_x \Delta_y \sum_{i=1}^{n_x} \sum_{j=1}^{n_y}
	\cos \left( k_x \Delta_x (i-1) \right)
	\cos \left( k_y \Delta_y (j-1) \right) f (x, y)  .
\end{equation}

We aim to determine which wavenumbers, $\textbf{k} = (k_x, k_y)$, can be eventually excited in the course of a simulation. Given a system with $R$ reactions and with two species, U and V, spread over $I$ compartments we have the chemical master equation (CME), as follows
\begin{equation}
	\frac{\p }{\p t} P \left( \textbf{W}, t | \textbf{W}_0, t_0 \right) =
	\sum_{i=1}^{2I} \sum_{j=1}^R
	\left[ P \left( \textbf{W} - \bm{\nu}_j, t | \textbf{W}_0, t_0 \right)
	a_{i,j} \left( \textbf{W} - \bm{\nu_j} \right) -
	P \left( \textbf{W}, t | \textbf{W}_0, t_0 \right)
	a_{i,j} \left( \textbf{W} \right) \right] .
	\label{Eq:CME}
\end{equation}
The vector $\textbf{W} = \left( \textbf{U}, \textbf{V} \right)$ is composed of
$\textbf{U} = \left( U_1, U_2, \ldots, U_I \right)^T$ and $\textbf{V} = \left( V_1, V_2, \ldots, V_I \right)^T$, which are vectors of the number of molecules in each compartment of species U and V, respectively. The function $a_{i,j}$ is the propensity function for reaction $j$ in compartment $i$, and $\nu_{ij}$ is the stoichiometric matrix. Now, we assume, that the number of molecules in compartment $\chi_i$ can be expressed as
\begin{equation}
	U_i = \phi_i \Theta + \eta_{ui} \sqrt{\Theta}
	\quad \text{and} \quad
	V_i = \psi_i \Theta + \eta_{vi} \sqrt{\Theta}  ,
	\label{Eq:WNE_new_variables}
\end{equation}
where $\phi_i$ and $\psi_i$ are the expected ratios of the number of molecules to the system parameter, $\Theta$, which is the magnitude of the smallest homogeneous steady state, and $\eta_{ui}$ and $\eta_{vi}$ are random variables that describe the noise in species U and V, respectively. As we change variables, we redefine the probability $P$ to
\begin{equation}
	P(\textbf{U}, \textbf{V}, t ) = \Pi ( \bm{\eta}_u, \bm{\eta}_v, t) \, ,
\end{equation}
which we can differentiate with respect to time, keeping the number of molecules
constant, to arrive at
\begin{eqnarray}
	\frac{\p P}{\p t} &=&
	\frac{\p \Pi}{\p t} +
	\sum_{i=1}^I
	\left(
	\frac{\D \eta_{ui}}{\D t} \frac{\p \Pi}{\p \eta_{ui}} +
	\frac{\D \eta_{vi}}{\D t} \frac{\p \Pi}{\p \eta_{vi}}
	\right) \\
	&=& \frac{\p \Pi}{\p t} -
	\sqrt{\Theta} \sum_{i=1}^I
	\left(
	\frac{\D \phi_i}{\D t} \frac{\p \Pi}{\p \eta_{ui}} +
	\frac{\D \psi_i}{\D t} \frac{\p \Pi}{\p \eta_{vi}}
	\right)  .
	\label{Eq:WNE_time_derivative}
\end{eqnarray}
To simplify further calculations, let us define new variables
\begin{equation}
	\bm{\zeta} = (\bm{\eta}_{u},
	\bm{\eta}_{v}) \quad \text{and} \quad \bm{\varphi} = (\bm{\phi}, \bm{\psi})  .
\end{equation}
Therefore, Equation \eqref{Eq:WNE_time_derivative} becomes
\begin{equation}
	\frac{\p P}{\p t} = \frac{\p \Pi}{\p t} -
	\sqrt{\Theta} \sum_{i=1}^{2I} \frac{d \varphi_i}{dt} \frac{\p \Pi}{\p \zeta_i}  .
\end{equation}

Before we proceed further, we need to make a distinction between the macroscopic propensity function, $\mathrm{a}_{i,j}$, and the microscopic propensity function, $a_{i,j}$. The macroscopic propensity function depends on the expected ratios of both species, $\mathrm{a}_{i,j} (\bm{\phi}, \bm{\psi})$, while the microscopic propensity function depends on the number of molecules of both species, $a_{i,j} (\textbf{U}, \textbf{V})$. This is important to note, as when we substitute Equation~\eqref{Eq:WNE_new_variables} into the microscopic propensity function, we have
\begin{eqnarray}
	a_{i,j} ( \textbf{W} - \nu_j ) &=&
	a_{i,j} (\bm{\phi} \Theta + \bm{\eta} \sqrt{\Theta} - \bm{\nu}_j ) \\
	&=& \Theta \mathrm{a}_{i,j} \left (\bm{\phi} + \frac{\bm{\eta}}{\sqrt{\Theta}} -
	\frac{\bm{\nu}_j}{\Theta} \right) + o \left( \frac{1}{\Theta} \right) \\
	&= &\Theta \left( \mathrm{a}_{i,j} \left( \bm{\phi} \right) +
	\frac{1}{\sqrt{\Theta}} \sum_{i=1}^I \frac{\p \mathrm{a}_{i,j}}{\p \phi_i} \eta_{ui}
	+ o \left( \frac{1}{\Theta} \right) \right)  .
\end{eqnarray}
If we substitute Equation~\eqref{Eq:WNE_new_variables} into the CME, Equation~\eqref{Eq:CME}, we have
\begin{eqnarray}
\nonumber
	\frac{\p \Pi}{\p t} -
	\sqrt{\Theta} \sum_{i=1}^{2I} \frac{d \varphi_i}{d t} \frac{\p \Pi}{\p \zeta_i} &=&
	\sum_{j=1}^J
	\left(-\sqrt{\Theta} \sum_{i=1}^{2I} \nu_{ij} \frac{\p}{\p \zeta_i} +
	\frac{1}{2} \sum_{i=1}^{2I} \sum_{l=1}^{2I}
	\nu_{ij} \nu_{lj} \frac{\p^2}{\p \zeta_i \zeta_l} \right) \\
	&& \quad \qquad \times
	\left( \mathrm{a}_{i,j} (\bm{\varphi}) + \frac{1}{\sqrt{\Theta}} \sum_{i=1}^{2I}
	\frac{\p \mathrm{a}_{i,j}}{\p \varphi_i} \right) \Pi (\bm{\zeta}) .
\end{eqnarray}
Next, we expand the above in powers of $\Theta$. The leading term, of order $\sqrt{\Theta}$, gives
\begin{equation}
	\sum_{i=1}^{2I} \frac{\p \varphi_i}{\p t} \frac{\p \Pi}{\p \zeta_i} =
	\sum_{j=1}^R \sum_{i=1}^{2I} \nu_{ij} \frac{\p \Pi}{\p \zeta_i} \mathrm{a}_{i,j}  ,
\end{equation}
which are satisfied by
\begin{equation}
	\frac{\p \varphi_i}{\p t} = \sum_{j=1}^R \nu_{ij} \mathrm{a}_{i,j} (\bm{\varphi})  ,
\end{equation}
i.e. the mean-field equations. The next terms in the expansion, terms of order $1$, give
\begin{eqnarray}
	\frac{\p \Pi}{\p t} &=& - \sum_{j=1}^R \left[ \sum_{i=1}^{2I} \sum_{l=1}^{2I}
	\nu_{ij} \frac{\p \mathrm{a}_{i,j}}{\p \varphi_i} \frac{\p (\zeta_l \Pi)}{\p \zeta_i} +
	\frac{1}{2} \sum_{i=1}^{2I} \sum_{l=1}^{2I} \mathrm{a}_{i,j} \nu_{ij} \nu_{lj}
	\frac{\p^2 \Pi}{\p \zeta_i \p \zeta_l} \right]  \\
	&=& - \sum_{i=1}^{2I} \sum_{l=1}^{2I} A_{il}
	\frac{\p (\zeta_l \Pi)}{\p \zeta_i} +
	\frac{1}{2} \sum_{i=1}^{2I} \sum_{l=1}^{2I} B_{il}
	\frac{\p^2 \Pi}{\p \zeta_i \p \zeta_l}  ,
\end{eqnarray}
which is the Fokker-Planck equation for the noise in the system, as stated in~\cite{woolley2011power}. The WNE allows us to separate the scales and consider the effects of weak noise within the system.

We wish to use this method to determine which wave modes can be excited when we will consider pattern formation in a later section. As such, we will need to find the discrete cosine transform of the Laplacian, as was done in~\cite{woolley2011stochastic}, but for a system in two spatial dimensions. This leads to the following expressions for the evolution of the Fourier transformed covariances
\begin{eqnarray}
	\left\langle \dot{\widehat{\eta}_{u k_x k_y} \widehat{\eta}_{u k_x k_y}} \right\rangle &=& 2 d_1 \left\langle \widehat{\eta}_{u k_x k_y} \widehat{\eta}_{u k_x k_y} \right\rangle + 2 b_0 \left\langle \widehat{\eta}_{u k_x k_y} \widehat{\eta}_{v k_x k_y} \right\rangle + v_1 \, , \label{Eq:WNE_ODEs_I} \\
	\left\langle \dot{\widehat{\eta}_{u k_x k_y} \widehat{\eta}_{v k_x k_y}} \right\rangle &=& c_0 \left\langle \widehat{\eta}_{u k_x k_y} \widehat{\eta}_{u k_x k_y} \right\rangle + (d_1 + d_2) \left\langle \widehat{\eta}_{u k_x k_y} \widehat{\eta}_{v k_x k_y} \right\rangle \nonumber \\ && \qquad + b_0 \left\langle \widehat{\eta}_{v k_x k_y} \widehat{\eta}_{v k_x k_y} \right\rangle + v_2 \, , \label{Eq:WNE_ODEs_II} \\
	\left\langle \dot{\widehat{\eta}_{u k_x k_y} \widehat{\eta}_{u k_x k_y}} \right\rangle &=& 2 d_2 \left\langle \widehat{\eta}_{v k_x k_y} \widehat{\eta}_{v  k_x k_y} \right\rangle + 2 c_0 \left\langle \widehat{\eta}_{u k_x k_y} \widehat{\eta}_{v k_x k_y} \right\rangle + v_3 \, . \label{Eq:WNE_ODEs_III}
\end{eqnarray}
If we assume that the system under study can be expressed as in Equation~\eqref{Eq:general_system_equations_I} and Equation~\eqref{Eq:general_system_equations_II}, then the constants in the above equations have the following form
\begin{eqnarray}
	d_1 &=& 2 \lambda_{u,1} (\cos (k_x \kappa h ) - 1) +
	2 \lambda_{u,2} (\cos (k_y h ) - 1) \nonumber \\ && \qquad +
	4 \lambda_{u,3} (\cos (k_x \kappa h ) \cos (k_y h ) - 1) + f_\phi  , \\
	d_2 &=& 2 \lambda_{v,1} (\cos (k_x \kappa h ) - 1) +
	2 \lambda_{v,2} (\cos (k_y h ) - 1) \nonumber \\ && \qquad+
	4 \lambda_{v,3} (\cos (k_x \kappa h ) \cos (k_y h ) - 1) + g_\psi  , \\
	b_0 &=& g_\phi  , \\
	c_0 &=& f_\psi  , \\
	v_1 &=& \frac{K A^2}{2} \left( \lambda_{u, 0} \bar{u} -
	2 \lambda_{u, 1} \bar{u} \cos (k_x \kappa h) -
	2 \lambda_{u, 3} \bar{u} \cos (k_y h) \right.  \\
	&&\left. \qquad  - 4 \lambda_{u, 2} \bar{u} \cos (k_x \kappa h) \cos (k_y h) +
	f(\bm{\phi}, \bm{\psi}) \right)  , \\
	v_2 &=& \frac{K A^2}{2} h(\bm{\phi}, \bm{\psi})  , \\
	v_3 &=& \frac{K A^2}{2} \left( \lambda_{v, 0} \bar{v} -
	2 \lambda_{v, 1} \bar{v} \cos (k_x \kappa h) -
	2 \lambda_{v, 3} \bar{v} \cos (k_y h)  \right.  \\
	&&\left. \qquad - 4 \lambda_{v, 2} \bar{v} \cos (k_x \kappa h) \cos (k_y h) +
	g(\bm{\phi}, \bm{\psi})  \right)  .
\end{eqnarray}
The function $h(\bm{\phi}, \bm{\psi})$ is composed of terms present in both functions $f$ and $g$, \textit{i.e.} the terms that represent reactions involving both species. The jump rates $\lambda_{u,i}$ and $\lambda_{v,i}$, are as defined in the main body of this paper. Equation~\eqref{Eq:WNE_ODEs_I}, Equation~\eqref{Eq:WNE_ODEs_II} and Equation~\eqref{Eq:WNE_ODEs_III} form a system of linear ordinary differential equations and, as a result, can be written
\begin{equation}
	\dot{\textbf{X}} = M \textbf{X} + V  , \label{Eq:WNE_matrix_equation}
\end{equation}
where $\textbf{X} = \left( \left\langle \widehat{\eta}_{u k_x k_y}
\widehat{\eta}_{u k_x k_y} \right\rangle,
\left\langle
\widehat{\eta}_{u k_x k_y}
\widehat{\eta}_{v k_x k_y}
\right\rangle,
\left\langle
\widehat{\eta}_{v k_x k_y}
\widehat{\eta}_{v k_x k_y}
\right\rangle\right)^T$,
$V = (v_1, v_2, v_3)^T$ and
\begin{equation}
	M =
\begin{bmatrix}
	2d_1 & 2b_0 & 0 \\
	c_0 & (d_1 + d_2) & b_0 \\
	0 & c_0 & 2d_2
\end{bmatrix}.
\end{equation}
Therefore, Equation~\eqref{Eq:WNE_matrix_equation} can be solved in the same manner as the system of ordinary differential Equations~\eqref{Eq:no_diffusion_ODE_system} and we have the following eigenvalues
\begin{eqnarray}
	e_0 &=& d_1 + d_2  , \\
	e_{\pm} &=& e_0 \pm \sqrt{4 b_0 c_0 + (d_1 - d_2)^2}  .
\end{eqnarray}
For spatial patterns to form, we require at least one of the eigenvalues to
have a positive real part. However, since $\cos(\cdot) - 1 \leq 0$ and the
condition for a Turing instability is $f_\phi + g_\psi < 0$, as stated in
Section \ref{Sec_supp_linear_stability_analysis}, then $d_1 + d_2 \leq 0$.
As a result, $\text{Re}(e_0) \leq 0$ and $\text{Re}(e_{-}) \leq 0$.
Therefore, the only way for the patterns to form is if
$\text{Re} (e_{+}) \geq 0$, and so the condition for patterns to form is
\begin{equation}
	d_1 d_2 - b_0 c_0 < 0 \, .
\end{equation}


\bibliography{refs}
\bibliographystyle{abbrv}
